\documentclass[a4paper,twocolumn,aps,eqsecnum,nofootinbib,notitlepage]{revtex4-1}
\usepackage{graphicx}% Include figure files
\usepackage{bm}% bold math
\usepackage{amsmath}
\usepackage{amssymb}
\usepackage{amsfonts}
\newcommand{\half}{\frac{1}{2}}

\newcommand{\der}{\partial}

\newcommand{\ket}{\rangle}

\newcommand{\dsl}{\partial\kern-0.55em\raise 0.14ex\hbox{/}}
\newcommand{\bfe}{\bm{e}}

\newcommand{\bfk}{\bm{k}}

\newcommand{\bfn}{\bm{n}}
\newcommand{\bfp}{\bm{p}}
\newcommand{\bfq}{\bm{q}}
\newcommand{\bfr}{\bm{r}}

\newcommand{\bfx}{\bm{x}}

\begin{document}

%\preprint{KYUSHU-HET-123}

\title{Numerical study of renormalization group flows of nuclear
effective field theory without pions on a lattice}

\author{Koji Harada}
\email{harada@artsci.kyushu-u.ac.jp}
\affiliation{Faculty of Arts and Science, Kyushu University\\
Fukuoka 819-0395 Japan}

\author{Satoru Sasabe}
\email{sasabe@phys.kyushu-u.ac.jp \ (corresponding author)}
\affiliation{Department of Physics, Kyushu University \\
Fukuoka 819-0395 Japan}

\author{Masanobu Yahiro}
\email{yahiro@phys.kyushu-u.ac.jp}
\affiliation{Department of Physics, Kyushu University \\
Fukuoka 819-0395 Japan}

\date{\today}

\begin{abstract}
 We formulate the next-to-leading order nuclear effective field theory
 without pions in the two-nucleon sector on a spatial lattice, and
 investigate nonperturbative renormalization group flows in the strong
 coupling region by diagonalizing the Hamiltonian numerically.  The
 cutoff (proportional to the inverse of the lattice constant)
 dependence of the coupling constants is obtained by changing the
 lattice constant with the binding energy and the asymptotic
 normalization constant for the groundstate being fixed.  We argue
 that the critical line can be obtained by looking at the finite-size
 dependence of the groundstate energy.  We determine the relevant
 operator and locate the nontrivial fixed point, as well as the
 physical flow line corresponding to the deuteron in the
 two-dimensional plane of dimensionless coupling constants. It turns
 out that the location of the nontrivial fixed point is very close to
 the one obtained by the corresponding analytic calculation, but the
 relevant operator is quite different.
\end{abstract}

% insert suggested PACS numbers in braces on next line
%\pacs{}
% insert suggested keywords - APS authors don't need to do this
%\keywords{}

\maketitle

%\iftrue
\iffalse
\newcommand{\comment}[1]{{\color{red}\tiny #1}}
\else
\newcommand{\comment}[1]{}
\fi

\section{Introduction}

\comment{NEFT}

Since the seminal work by Weinberg~\cite{Weinberg:1990rz,
Weinberg:1991um, Weinberg:1992yk}, the low-energy effective field
theory of nucleons (and other low-energy excitations, such as pions),
the so-called nuclear effective field theory (NEFT), has been
investigated extensively; see Refs.~\cite{Epelbaum:2008ga,
Machleidt:2011zz} for the reviews. In NEFT, the ``fundamental''
degrees of freedom are low-lying hadrons so that NEFT is applicable
only up to a certain momentum scale, the physical cutoff
$\Lambda_\text{phys}$.  The effects of heavier degrees of freedom than
$\Lambda_\text{phys}$, the processes with momenta higher than
$\Lambda_\text{phys}$, and the internal structure of the hadrons are
integrated out and have been encoded in the coupling constants of
local interactions.  For example the effects of heavy-meson exchange
processes between two nucleons are represented by four-nucleon
operators. Note that even if pions are included in NEFT, the exchange
of the pion with momentum transfer higher than the cutoff is
represented as local four-nucleon (and $2n$-nucleon, in general)
operators.

\comment{NEFT on a lattice}

Although the early investigations exclusively employed continuous,
semi-analytic approach based on the Lippmann-Schwinger (LS) equation,
the Faddeev equation, etc., the methods of numerical simulation on a
lattice have been developed recently~\cite{Lee:2004si, Borasoy:2006qn,
Borasoy:2007vi, Abe:2007fe, Epelbaum:2009zsa, Lahde:2013uqa,
Epelbaum:2013paa, Lynn:2014zia, Lahde:2015ona, Wlazlowski:2014jna,
Gezerlis:2013ipa, Gezerlis:2014zia, Tews:2015ufa, Klein:2015vna}; see
Ref.~\cite{Lee:2008fa} for the review. The inverse of the lattice
constant provides the cutoff in momentum. It should not exceed the
physical cutoff. Note that, unlike lattice QCD, we should not take the
continuum limit in lattice NEFT.

\comment{advantages}

Lattice simulation of NEFT is very interesting since it has several
advantages. First of all, it allows us to calculate many-nucleon
quantities without suffering from complications due to the increase of
the number of nucleons. Remember that the Faddeev equation for three
nucleons is more complex than the two-nucleon LS equation, and the
Faddeev-Yakubovsky equation for four nucleons is even more
complex. Lattice formulation does not have this kind of complication,
except for the construction of many-nucleon operators,
which are necessary when the correlators are calculated.
By now, considerably large nuclei have already been investigated on a
lattice~\cite{Epelbaum:2011md, Lahde:2013uqa,
Epelbaum:2013paa}. Second, arbitrarily complicated pion interactions
can be included in lattice simulations, just as arbitrarily
complicated gluon interactions can be included in lattice QCD. It thus
provides the possibility of the calculations with the \textit{exactly
chiral symmetric} interactions of pions with nucleons that are
nonlinearly realized. This direction of investigation is now in
progress~\cite{HNSY}. Note that the truncation of pion interactions at
a finite order inevitably breaks chiral symmetry. Third, it is
straightforward to make the system contact to a heat reservoir, and
also to a particle reservoir.

\comment{why RG?}

In order to perform lattice simulation of NEFT effectively, it is
essential to understand how the operators behave on a lattice. That
is, we need to know how the coupling constants depend on the lattice
constant, hence on the momentum cutoff. This is a typical
renormalization group (RG) problem.

\comment{finely tuned}

It is critically important that the two-nucleon systems in the S waves
are finely tuned: the scattering lengths are \textit{unnaturally}
long. From the RG point of view, this may be viewed as an evidence of
the fact that the system is near a nontrivial fixed point. The
nontrivial fixed point is located on a critical surface, which is the
boundary between the strong and weak coupling phases. The scattering
length would be infinite if the system were on the critical surface.

\comment{anomalous dimensions and the relevant operator}

Near the nontrivial fixed point, operators gain anomalous dimensions
and hence their behavior is quite different from that in the
perturbative region near the trivial fixed point where all the
interactions are turned off.  Anomalous dimensions change the
importance of the operators. It has been shown that there is a
relevant operator near the nontrivial fixed
point~\cite{Weinberg:1991um}. (Remember that relevant operators are
the ones with positive scaling dimensions and the most important at
low energies.) The RG flow tells us what the relevant operator is.
Pictorially it is the direction of the flow going out from the
nontrivial fixed point. A good example is the dashed line in
Fig.~\ref{combined5} shown later.

\comment{the location of the nontrivial fixed point and the flow are
 not universal}

While the scaling dimensions are universal, the location of the
nontrivial fixed point is not, i.e., it depends on how the
cutoff is implemented. The scaling dimensions are obtained in the
literature~\cite{Birse:1998dk, Harada:2005tw, Harada:2006cw} by using
a continuum formulation, and they must be the same for lattice
regularization. The RG flow and hence the relevant operator are not
universal. They on a lattice should be determined by explicit
calculations.  It is our purpose of the present paper to numerically
determine the location of the nontrivial fixed point and the relevant
operator of NEFT without pions on a lattice.

\comment{the physical importance}

The determination of the location of the nontrivial fixed point and
the RG flow are of direct phyical importance. From the information, we
know the relevant operator that dominates physics. It is also
important to know which flow line corresponds to the physical system
in order to perform numerical simulations since it provides the input
parameters.

\comment{inclusion of pions does not alter the result}

Even if pions are included, the results do not drastically change from
those without pions. The strong short-distance part of pion exchange
interactions is cutoff on a lattice. See Ref.~\cite{Harada:2010ba} for
the effects of pions in the EFT with a finite cutoff. The study here
therefore is an important step toward the chirally-symmetric NEFT with
pions on a lattice.

\comment{Seki-van Kolck}

NEFT without pions with a lattice regularization has been considered
by Seki and van Kolck~\cite{Seki:2005ns} based on the analytic
approach. Starting with the continuum S-wave LS equation, they replace
the momentum integrals with the ones over the first Brillouin zone and
the momentum squares in the integrands with the corresponding
discretized expressions to imitate the theory defined on a lattice,
and determine the dependence of the scattering length and the
effective range on the lattice constant.

\comment{Seki-van Kolck does not yield a genuine lattice result}

Note that the method of Seki and van Kolck does not yield a genuine
lattice result. Theory defined on a lattice has explicit rotational
invariance breaking so that the amplitude is an admixture of ``partial
waves.'' On the other hand, the starting point of Seki and van Kolck
is the LS equation in a specific partial wave, derived in the
continuum theory.

\comment{the method: hamiltonian diagonalization}

In this paper, we employ the numerical diagonalization of the NLO
Hamiltonian of NEFT without pions defined on a spatial cubic lattice
with periodic boundary condition in order to investigate RG flows in
the strong coupling phase where a single two-nucleon boundstate
appears as the groundstate.  Of course, the RG flows may be obtained
by using a Monte-Carlo simulation, but the diagonalization is much
more accurate and numerically simpler. We confine ourselves to the
strong coupling phase because in the weak coupling phase, where
boundstates are absent physically, but the groundstate is found to
have negative energy due to the periodic boundary
condition~\cite{Luscher:1986pf}.

\comment{finite-size effect on the groundstate energy}

We argue that at the phase boundary the finite-size effect is
maximal. By looking at the finite-size dependence of the groundstate
energy, we can infer the location of the phase boundary. It turns out
that, with this information on the phase boundary, the location of the
nontrivial fixed point can be determined quite accurately. We
emphasize that this is the first determination of the location of the
nontrivial fixed point and the RG flow of the NEFT genuinely defined
on a lattice.

\comment{improvement of the discretization}

We consider the improvement of the discretized representation of the
derivatives to reduce the finite lattice constant errors and the
rotational symmetry breaking effects.

\comment{comparison with the analytic method}

We compare the numerical results with those obtained by the analytic
method with a lattice regularization, which is a generalization of the
method of Seki and van Kolck. They treat the NLO coupling
perturbatively, but we deal with it nonperturbatively by considering
full dependence of the couplings on the scattering amplitude.

\comment{structure of the paper}

The structure of the paper is the following: In Sec.~\ref{sec:setup},
we recapitulate the continuum approach to the NLO NEFT. The RG
equations and the flow~\cite{Harada:2005tw} are obtained by solving
the LS equation for the scattering amplitude.  We also consider the
Schr\"odinger equation and the boundstate, and show that the
asymptotic normalization constant (ANC)~\cite{ANC} can be used as
a low-energy physical constant. In Sec.~\ref{sec:analytic} we evaluate
the integrals appearing in the analysis by using the lattice
regularization in the same spirit of Seki and van
Kolck~\cite{Seki:2005ns}, and draw the RG flows with and without the
improvement of the discretization of the derivatives. We then switch
to the numerical diagonalization of the Hamiltonian defined on a
lattice and obtain the RG flow in the strong coupling phase by
requiring the binding energy and the ANC to be independent of the
cutoff in Sec.~\ref{sec:lattice}. We consider the lattice finite-size
effect of the groundstate energy and argue that it is maximal at the
phase boundary. With this information, together with the RG flow, we can
determine the location of the nontrivial fixed point. In
Sec.~\ref{sec:summary}, the summary of the results and the discussions
are given. In Appendix~\ref{sec:constants}, we outline the evaluation
of the integral that is necessary to draw the RG flow with the
improved discretization.

\section{NLO NEFT without pions}
\label{sec:setup}

\subsection{Renormalization group equations and the flow}

\comment{setup}

In order to set up an effective field theory, one needs to choose the
relevant degrees of freedom and the accuracy to be achieved. In the
NEFT without pions, we consider only nucleons. There are an infinite
number of local operators that represents interactions among
nucleons. The leading order (LO) interaction is represented by the
momentum independent four-nucleon operator, and the next-to-leading
order (NLO) one by the four-nucleon operator with two
spatial derivatives. In the following, we will concentrate on the
S waves.

\comment{Lagrangian}

The isospin SU(2) symmetric Lagrangian of our effective field theory
is given by
\begin{align}
  \mathcal{L} &= N^\dagger
  \left(i\der_t+\frac{\nabla^2}{2M}\right)N-C_0 (N^TP_k N)^\dagger
  (N^T P_k N)
  \notag \\
 &{}\quad +C_2\left[
 (N^TP_k N)^\dagger (N^T P_k\overleftrightarrow{\nabla}^2 N) + \text{h.c.}
 \right],
  \label{lag}
\end{align}
where $N$ is the nucleon operator, $M$ represents the mass, and
$\overleftrightarrow{\nabla}^2 = \overleftarrow{\nabla} \cdot
\overleftarrow{\nabla} - 2\overleftarrow{\nabla} \cdot
\overrightarrow{\nabla} + \overrightarrow{\nabla} \cdot
\overrightarrow{\nabla}$.  The terms higher than NLO are
omitted. $P_k$ is a projection operator; for the $^3S_1$ (spin
triplet) channel, it is $P_k=\sigma^2 \sigma^k \tau^2/\sqrt{8}$, with
$\sigma^a$ and $\tau^a$ being spin and isospin Pauli matrices
respectively.

\comment{LS eq.}

The LS equation for the off-shell center-of-mass nucleon-nucleon (NN)
scattering amplitude is given by
\begin{eqnarray}
 &&\!\!\!-i{\cal A}(p^0,\bfp_1, \bfp_2)=-iV(\bfp_1, \bfp_2) \nonumber \\
 &&\!\!\!{}+\!\!\int \!\!\!\frac{d^3k}{(2\pi)^3}
  \left(-iV(\bfk,\bfp_2)\right)\!
  \frac{i}{p^0\!-\!\bfk^2/M\!+\!i\epsilon}
  \left(\!-i {\cal A}(p^0,\bfp_1,\bfk)\right), \nonumber \\
\end{eqnarray}
where $V$ is the vertex in momentum space,
\begin{equation}
 V(\bfp_1,\bfp_2)=C_0+4C_2\left(\bfp_1^2+\bfp_2^2\right),
\end{equation}
and $p^0$ is the (off-shell) center-of-mass energy of the system,
$\bfp_1$ and $\bfp_2$ are half the relative momenta in the
initial and final two-nucleon states respectively.

\comment{solution}

The solution of this LS equation is obtained\cite{Gegelia:1998iu,
Phillips:1997xu,Harada:2005tw} as
\begin{equation}
 {\cal A}(p^0,\bfp_1,\bfp_2)=x(p^0) 
  + y(p^0)(\bfp_1^2+\bfp_2^2)+z(p^0) \bfp_1^2\bfp_2^2,
\end{equation}
with
\begin{align}
 x&=\left(C_0+16C_2^2I_2\right)/D, \\
 y&=4C_2\left(1-4C_2I_1\right)/D, \\
 z&=16C_2^2I_0/D,
\end{align}
where we have introduced
\begin{equation}
 D=1-C_0I_0-8C_2I_1+16C_2^2I_1^2-16C_2^2I_0I_2\, ,
  \label{denom}
\end{equation}
and
\begin{equation}
 I_n=-M\int \frac{d^3 k}{(2\pi)^3}
  \frac{|\bfk|^{2n}}{|\bfk|^2+\mu^2}\, ,
  \quad \mu=\sqrt{-Mp^0-i\epsilon}\, .
\end{equation}
The integrals $I_n$ are divergent and require regularization. If we
impose a sharp momentum cutoff $\Lambda$, they are given as
\begin{equation}
 I_n=-\frac{M}{2\pi^2}\int_0^\Lambda dk \frac{k^{2n+2}}{k^2+\mu^2}\, ,
  \quad \mu=\sqrt{-Mp^0-i\epsilon}\, .
\end{equation}

\comment{RG analysis}

The Wilsonian RG analysis of this system is formulated elegantly by
introducing the (energy-dependent) redundant operators, which can be
eliminated by the use of equations of
motion~\cite{Birse:1998dk,Harada:2005tw,Harada:2006cw}. However, for
our present purpose, it is simpler to consider the on-shell
formulation: we require the scattering length $a_0$ and the effective
range $r_0$ to be independent of the cutoff. See
Ref.~\cite{Harada:2005tw} for the relation between the two
formulations. At low energies, the on-shell amplitude can be written
as
\begin{align}
 \left.{\cal A}^{-1}\right|_\text{on-shell}&=
  -\frac{M}{4\pi}
  \left[
   -\frac{1}{a_0}+\half r_0 p^2 +{\cal O}(p^4)-ip
  \right], \notag \\
 &\text{with}\ 
  p=\sqrt{Mp^0}=\left|\bfp_1\right|=\left|\bfp_2\right|.
\end{align}
The scattering length and the effective range are given by
\begin{align}
 \frac{M}{4\pi}\frac{1}{a_0} &= \frac{M\Lambda}{2\pi^2}
  \left[
   \theta_1+\frac{(1+\theta_3Y)^2}{X-\theta_5Y^2}
  \right],
  \label{a0}
  \\
 \frac{M}{4\pi}\frac{r_0}{2} &= \frac{M}{2\pi^2\Lambda}
  \left[
   -R(0)+\frac{Y(2+\theta_3Y)(1+\theta_3Y)^2}{(X-\theta_5Y^2)^2}
  \right],
  \label{r0}
\end{align}
% \begin{align}
%  \frac{M}{4\pi}\frac{1}{a_0} &= \frac{M\Lambda}{2\pi^2}
%   \left[
%    1+\frac{(1+Y/3)^2}{X-Y^2/5}
%   \right],
%   \label{HIKa0}
%   \\
%  \frac{M}{4\pi}\frac{r_0}{2} &= \frac{M}{2\pi^2\Lambda}
%   \left[
%    1+\frac{Y(2+Y/3)(1+Y/3)^2}{(X-Y^2/5)^2}
%   \right],
%   \label{HIKr0}
% \end{align}
where we have introduced dimensionless coupling constants $X$ and $Y$
defined by
\begin{equation}
 C_0=\frac{2\pi^2}{M\Lambda}X, \quad 4C_2=\frac{2\pi^2}{M\Lambda^3}Y,
  \label{c0c2}
\end{equation}
as well as the constants $\theta_n \ (n=1,3,5)$ and the function
$R(x)$ defined by
\begin{align}
 I_0 &= -\frac{M\Lambda}{2\pi^2}
  \left[
   \theta_1
   +
   \left(\frac{p^2}{\Lambda^2}\right)
   R\left(\frac{p^2}{\Lambda^2}\right)
 \right]-\frac{iM}{4\pi}p\,,
  \label{I0}\\
 L_3&\equiv -M\int \frac{d^3k}{(2\pi)^3} =
 -\frac{M\Lambda^3}{2\pi^2}\theta_3\,,
 \label{L3} \\
 L_5&\equiv -M\int \frac{d^3k}{(2\pi)^3}|\bfk|^2 =
  -\frac{M\Lambda^5}{2\pi^2}\theta_5\,,
 \label{L5}
\end{align}
according to Seki and van Kolck~\cite{Seki:2005ns}. For the
regularization with the sharp momentum cutoff $\Lambda$, we have
\begin{equation}
 \theta_1=1\,, \quad \theta_3=\frac{1}{3}\,, \quad \theta_5=\frac{1}{5}\,,
  \quad
  R(0)=-1\, .
\end{equation}

\comment{Seki and van Kolck}

Note that Seki and van Kolck disregard the terms higher than linear in
$Y$ of Eqs.~\eqref{a0} and \eqref{r0}.

\comment{RGEs}

\begin{figure}[h]
 \includegraphics[width=0.9\linewidth,clip]{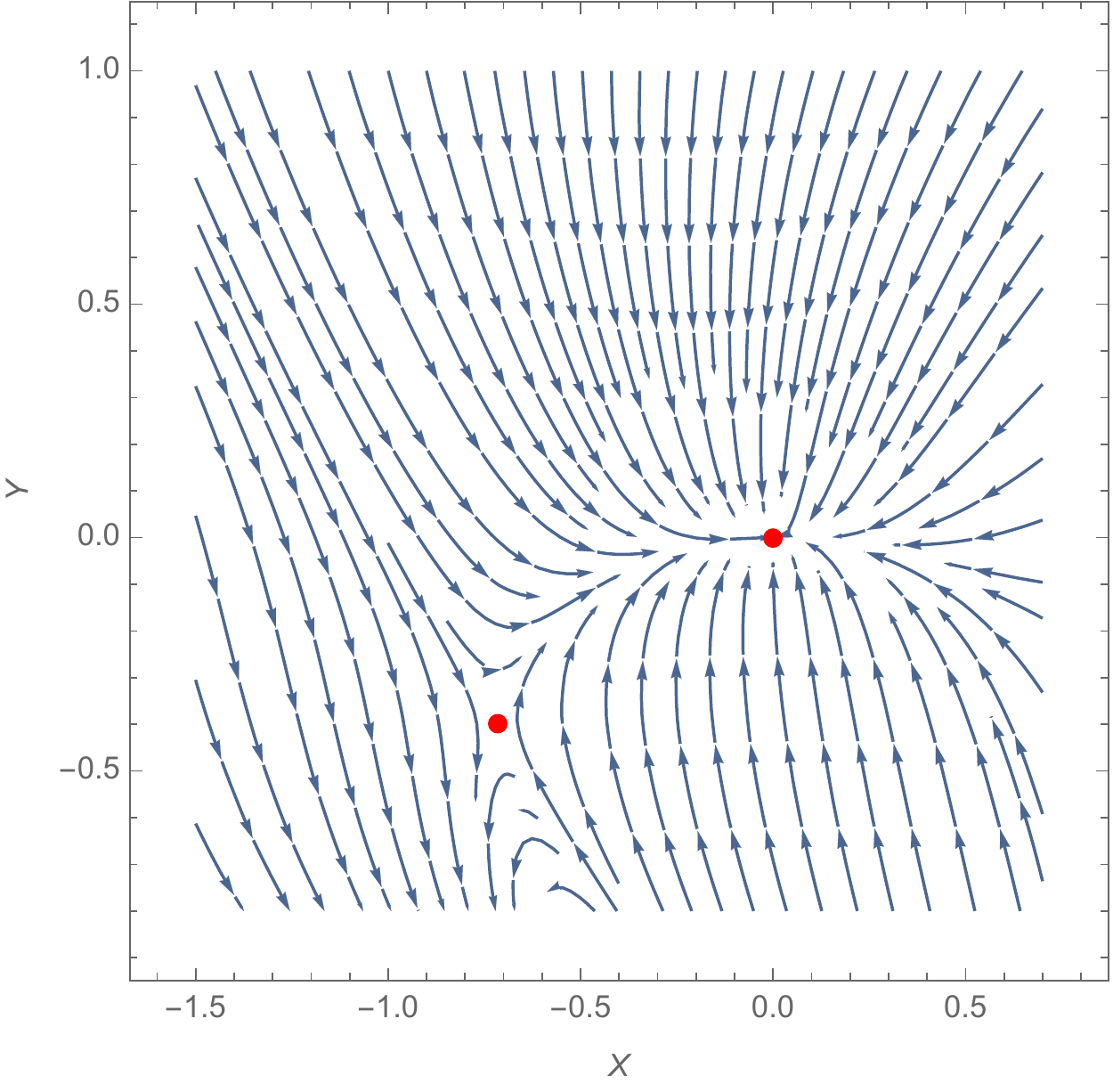}
 \caption{\label{continuumflow} The flow and the fixed points of the
 NLO NEFT in the $X$-$Y$ plane obtained by using a sharp momentum
 cutoff in the continuum formulation. The arrows indicate the
 directions of the smaller values of the cutoff.}
\end{figure}

By requiring that $a_0$ and $r_0$ are independent of $\Lambda$, we
obtain the following RG equations:
\begin{widetext}
  \begin{align}
  \Lambda\frac{dX}{d\Lambda} &=
   X\left(1+6\theta_3Y\right)
   +Y^2(5\theta_5+3\theta_3^2 X +3\theta_3\theta_5 Y)
   \notag \\
  &{}\quad+\frac{X-\theta_5 Y^2}{(1+\theta_3 Y)^2}
   \left[
    -R(0)(\theta_3 X+\theta_5Y)(X-\theta_5 Y^2)
    +\theta_1\left\{
              \theta_5Y^2(3+2\theta_3 Y)
              +X(1+2\theta_3 Y(2+\theta_3Y))
             \right\}
   \right], \label{rgeX}\\
  \Lambda\frac{dY}{d\Lambda} &=
   3Y\left(1+\frac{\theta_3}{2}Y\right)(1+\theta_3 Y)
   +\frac{X-\theta_5 Y^2}{2(1+\theta_3Y)}
   \left[
    -R(0)X+4\theta_1 Y +(R(0)\theta_5+2\theta_1\theta_3)Y^2
   \right]. \label{rgeY}
 \end{align}
 % \begin{align}
 %  \Lambda\frac{dX}{d\Lambda} &=
 %   X(1+2Y)+Y^2\left(1+\frac{X}{3}+\frac{Y}{5}\right)
 %   +\frac{X-Y^2/5}{\left(1+Y/3\right)^2}
 %   \left[
 %    X-\frac{Y^2}{5}+\left(\frac{X}{3}+\frac{Y}{5}\right)
 %    \left(X+4Y+\frac{7}{15}Y^2\right)
 %   \right],
 %   \label{HIKRG1}\\
 %  \Lambda\frac{dY}{d\Lambda} &=
 %   3Y\left(1+\frac{Y}{6}\right)\left(1+\frac{Y}{3}\right)
 %   +\frac{X-Y^2/5}{2\left(1+Y/3\right)}
 %   \left(X+4Y+\frac{7}{15}Y^2\right).
 %   \label{HIKRG2}
 % \end{align}
\end{widetext}
From these RG equations, we obtain the nontrivial fixed point as
\begin{align}
 (X_\star,Y_\star)
 &=\left(\frac{3}{5}(4-3\sqrt{3}),\frac{3}{2}(-2+\sqrt{3})\right)
  \notag \\
 &=(-0.717691\ldots,-0.401924\ldots),
  \label{physicalFP}
\end{align}
which is responsible for the ``unnaturally'' large scattering length
in the $^3S_1$ channel. The flow and the nontrivial fixed point (as
well as the trivial one) is depicted in Fig.~\ref{continuumflow}.

\comment{universal v.s. nonuniversal}

Although the existence of the nontrivial fixed point and the scaling
dimensions, which are the eigenvalues of the linearized RG equations
in the vicinity of the nontrivial fixed point, are universal, the
location of the fixed point and the flow are not universal: they
depend on the details of the regularization. In the following
sections, we will see how they vary as the regularization is changed.

\subsection{Groundstate wavefunction and the ANC}

\comment{Schr\"odinger eq.}

For the comparison with the results obtained in
Sec.~\ref{sec:lattice}, let us consider the (stationary) Schr\"odinger
equation for the relative motion of the two-nucleon boundstate in
momentum space,
\begin{align}
 E\psi(\bfp)&=
  \frac{\bfp^2}{M}\psi(\bfp)
 % \notag \\
 % &{}\quad
  +\int^{\Lambda} \frac{d^3q}{(2\pi)^3}\left[
 % \notag \\
 % &{}\qquad\qquad\quad\!
 C_0+4C_2(\bfp^2+\bfq^2)\right]\psi(\bfq),
 \label{continuumSchroedinger}
\end{align}
where $\psi(\bfp)$ is the cutoff wavefunction satisfying
$\psi(\bfp)=0$ for $|\bfp|>\Lambda$ and $E$ is the (negative) energy
eigenvalue. The integration is over the region $|\bfq|\le \Lambda$.

\comment{solution}

The Schr\"odinger equation can be solved as
\begin{equation}
 \psi(\bfp) =\frac{-M}{\bfp^2+\mu^2}
  \left[
   (C_0+4C_2\bfp^2)\alpha+4C_2\beta
  \right],
  \label{wf}
\end{equation}
where $\mu=\sqrt{M|E|}$, and $\alpha$ and $\beta$ are constants
defined by
\begin{equation}
 \alpha=\int^{\Lambda} \frac{d^3q}{(2\pi)^3} \psi(\bfq), \quad
 \beta=\int^{\Lambda} \frac{d^3q}{(2\pi)^3} \bfq^2 \psi(\bfq).
\end{equation}
By multiplying Eq.~\eqref{wf} by $(2\pi)^{-3}$ and
$\bfp^2(2\pi)^{-3}$, and integrating over $\bfp$, we obtain
\begin{align}
 \alpha&= (C_0\alpha+4C_2\beta)I_0+4C_2\alpha I_1, \label{alpha}\\
 \beta &= (C_0\alpha+4C_2\beta)I_1+4C_2\alpha I_2, \label{beta}
\end{align}
respectively. These equations have a nonzero solution if 
\begin{equation}
 \left|
    \begin{array}{cc}
     1-C_0I_0-4C_2I_1& -4C_2I_0\\
     -C_0I_1-4C_2I_2& 1-4C_2I_1\\
    \end{array}
   \right|=0. \label{det}
\end{equation}
This determinant is equal to $D$ given in Eq.~\eqref{denom}.  This
condition (the vanishing of the denominator of the scattering
amplitude) determines $\mu$, hence the energy eigenvalue $E$, and the
ratio $\beta/\alpha$,
\begin{equation}
 \frac{\beta}{\alpha}=\frac{1-C_0I_0-4C_2I_1}{4C_2I_0}.
\end{equation}

\comment{wavefunction}

The wavefunction is written as
\begin{equation}
 \psi(\bfp)=A+\frac{B}{\bfp^2+\mu^2},
  \label{continuumwf}
\end{equation}
where
\begin{align}
 A&=-4M\alpha C_2, \\
 B&=4MC_2(\mu^2\alpha-\beta)-M\alpha C_0
 \notag \\
 &=M\alpha\left[4C_2\mu^2-(1-4C_2I_1)/I_0\right].
\end{align}
Note that, after Fourier transformation, the coordinate-space
wavefunction is written as a sum of the regularized delta function and
the regularized Yukawa function.

The overall normalization is determined by the condition $\int
d^3p/(2\pi)^3\, |\psi(\bfp)|^2=1$.

It is natural to define the ANC as $B/4\pi$, since the Yukawa function
governs the asymptotic behavior of the wave function in the limit of
$\Lambda\to \infty$.

\comment{ANC as LEC}

In Sec.~\ref{sec:lattice}, we take the binding energy and the ANC as
low-energy physical quantities to be fixed to obtain the RG flow,
instead of the scattering length and the effective range. We will
numerically show that fixing the binding energy and the ANC is
equivalent to fixing the scattering length and the effective range.

\begin{figure}[h]
 \includegraphics[width=0.9\linewidth,clip]{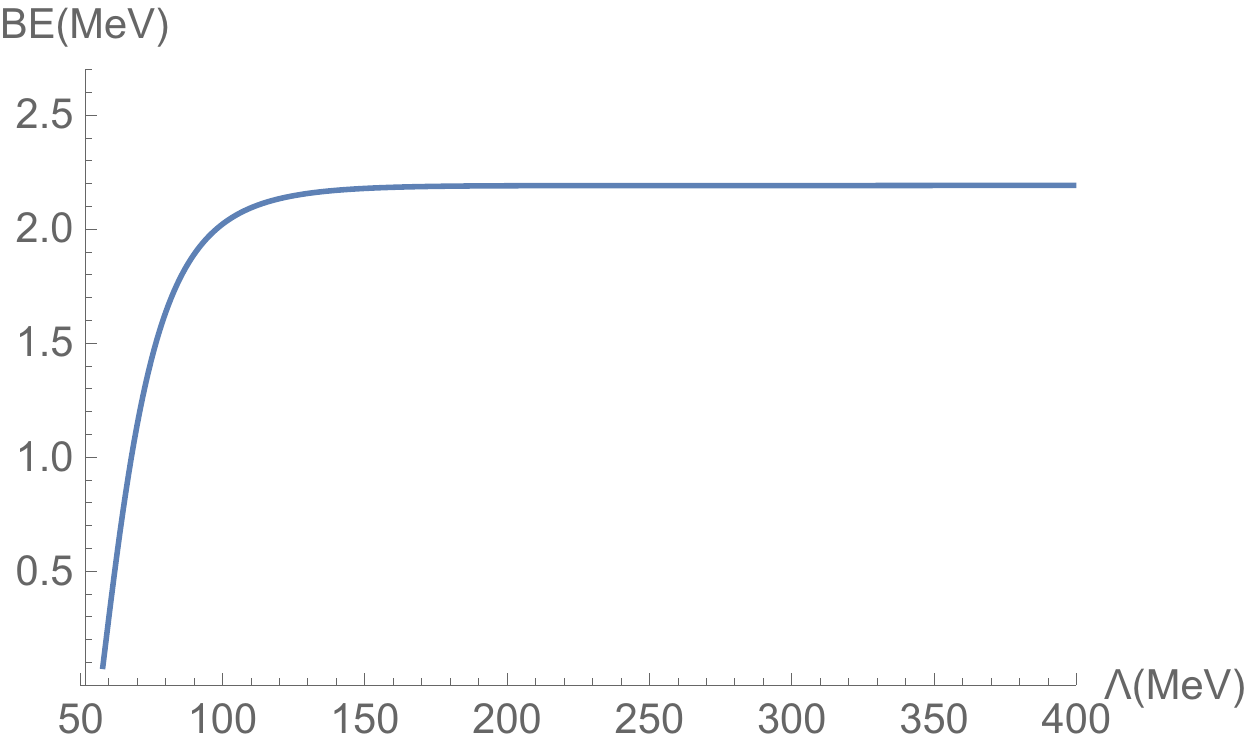}
 \includegraphics[width=0.9\linewidth,clip]{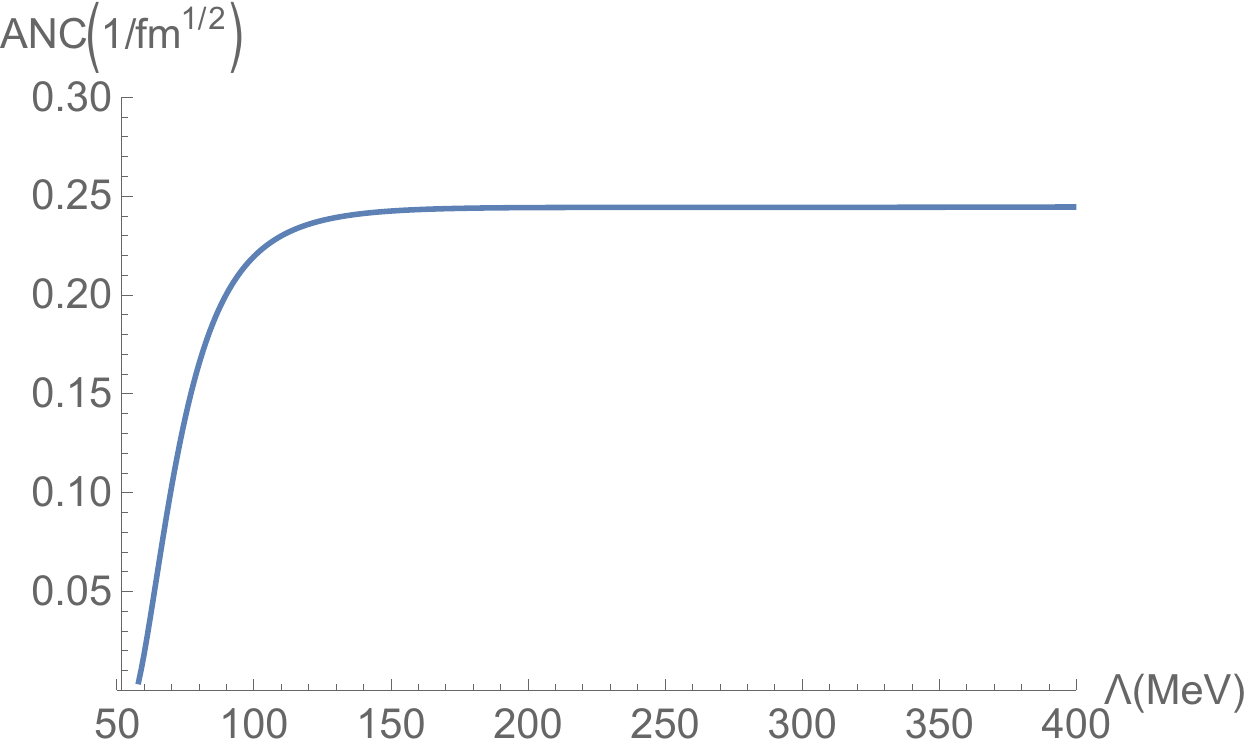}
 \caption{\label{be-anc} The cutoff dependence of the binding energy
 and the ANC, calculated for a given set of the scattering length and
 the effective range, $(a_0, r_0)=(5.42, 1.75)$ fm, corresponding to
 deuteron~\cite{deSwart:1995ui}.}
\end{figure}

Given the values of the scattering length and the effective range, we
determine the coupling constants $X$ and $Y$ by solving
Eqs.~\eqref{a0} and \eqref{r0} for each value of the cutoff
$\Lambda$. Then, solving Eq.~\eqref{det} numerically, we obtain the
binding energy and the ANC. The results for the deuteron scattering
length and the effective range are shown in Fig.~\ref{be-anc}. We see
that the binding energy and the ANC are constant (approximately equal
to 2.19 MeV and 0.244 fm$^{-1/2}$ respectively) for a wide range of
the cutoff.

\comment{lower cutoff, lower resolution}

Note also that both the binding energy and the ANC vanish at
$\Lambda\approx 57.2$ MeV, corresponding to 3.4 fm in length scale, or
$\Lambda^2/M\approx$ 3.5 MeV in energy scale. Physically it means that
the resolution there is too low to see the deuteron. Remember that
deuteron binding energy is 2.22 MeV, and the mean-square radius is
1.97 fm.

\comment{ANC is potential independent}

Incidentally, it is interesting to note that the value of the ANC we
obtain in our NLO NEFT, $B/4\pi=0.244$ fm$^{-1/2}$, is very close to
the recommended value in Ref.~\cite{deSwart:1995ui},
$0.8845(8)/\sqrt{4\pi}$ fm$^{-1/2}$ $= 0.2495(2)$ fm$^{-1/2}$,
obtained with a completely different NN potential. (The factor
$\sqrt{4\pi}$ comes from the normalization of the spherical
harmonics.)

\section{Analytic results with the lattice-regularized integrals}
\label{sec:analytic}

\comment{lattice regularization}

In this section, we consider the lattice regularization of the
integrals, following Seki and van Kolck~\cite{Seki:2005ns}. Consider
an infinitely large lattice with a finite lattice constant $a$. We
replace the integrals $I_n$ by the corresponding ones integrated over
the first Brillouin zone,
\begin{equation}
 -\frac{\pi}{a} \le k_i \le \frac{\pi}{a} \quad  (i=1,2,3),
\end{equation}
and the momentum square $|\bfk|^2$ coming from the Laplacian
$\nabla^2$ in the continuum by the corresponding
discretized one from the finite difference representation. The
three-point formula corresponding to the replacement,
\begin{equation}
 |\bfk|^2 \to \frac{4}{a^2}\sum_{i=1}^3 \sin^2\left(\frac{k_ia}{2}\right),
\end{equation}
is widely used. We also consider the five-point formula,
\begin{equation}
 |\bfk|^2 \to \frac{4}{a^2}\sum_{i=1}^{3}
  \left[
   \sin^2\left(\frac{k_i a}{2}\right)
   +\frac{1}{3}\sin^4\left(\frac{k_i a}{2}\right)
  \right],
\end{equation}
which has higher-order discretization errors than the three-point formula.
Note that we use the same difference formula for both the interaction
term and the kinetic term. With the three-point formula, for example,
the integral $I_0$ is defined by
\begin{equation}
 I_0=\frac{M}{a}\prod_{i=1}^{3}\left[\int_{-\pi}^{\pi}\frac{dk_i}{2\pi}\right]
  \frac{1}{p^2-4\sum_{i=1}^{3}\sin^2(k_i/2)+i\epsilon}\, ,
  \label{watson}
\end{equation}
where the change of variables $k_i\to k_i/a$ has been performed so
that the integration variables are now dimensionless. We have also
introduced a dimensionless quantity $p=\sqrt{(Ma)(p^0a)}$.

\comment{not a genuine lattice result}

It is important to note that the prescription described above does not
produce a genuine lattice result. On a lattice the rotational
invariance is explicitly broken so that the notion of ``partial
waves'' is not good. In the above procedure, however, we start with
the continuum, rotational invariant theory, derive the LS equation for
the S waves, solve it formally without specifying the regularization
of the integrals, and finally invoke the lattice
regularization. Although this prescription is not fully consistent,
the analytic results are a very useful guide for the genuine lattice
study, as shown later in Sec.~\ref{sec:lattRG}.

\comment{seki-van kolck did it}

\begin{figure}[h]
 \includegraphics[width=0.9\linewidth,clip]{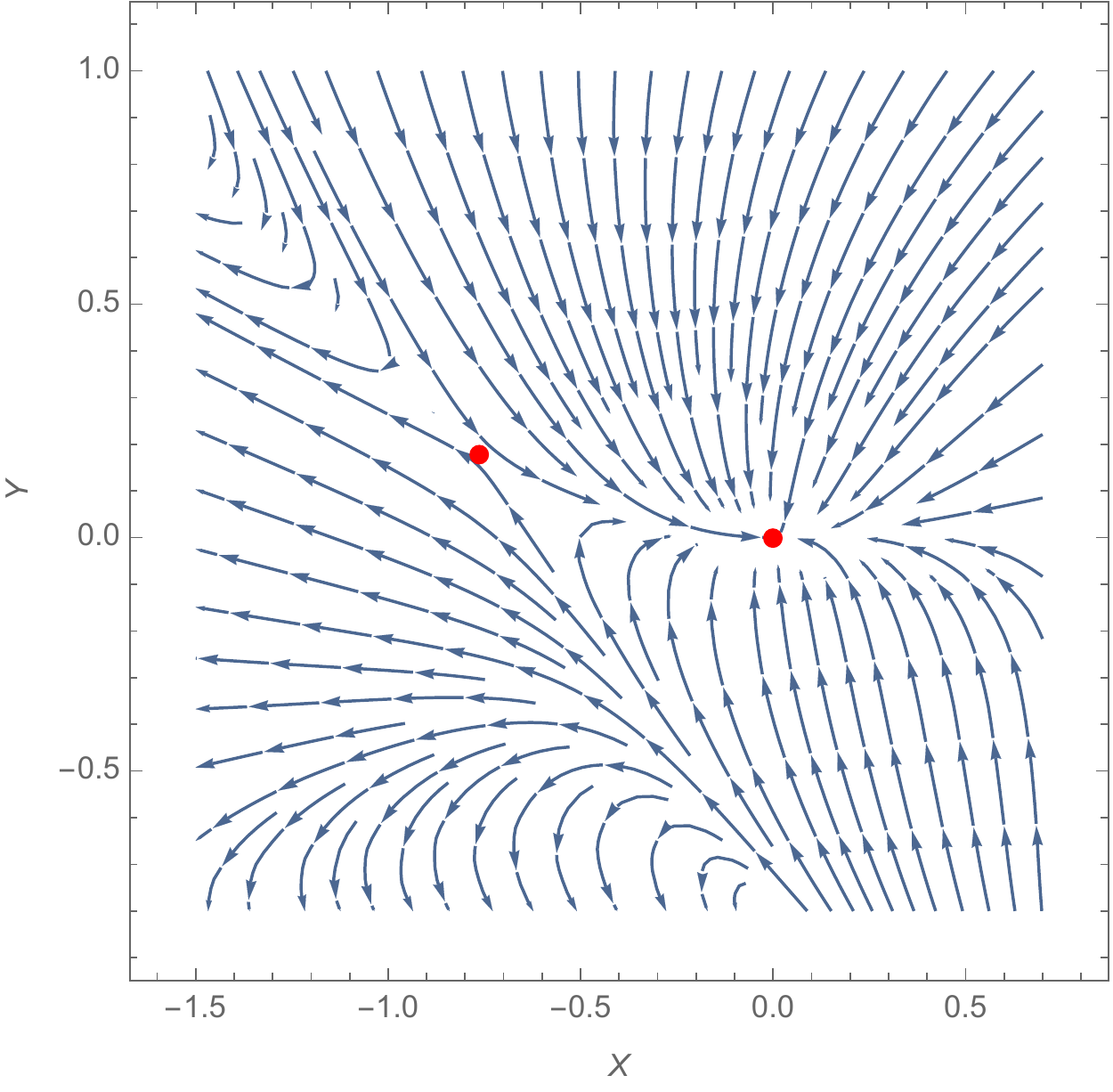}
 \caption{\label{lattice_3pt_flow} The flow and the fixed points of the
 NLO NEFT in the $X$-$Y$ plane obtained by using a lattice
 regularization with the 3-point formula.
 % The allows indicate the directions of
 % the smaller values of the cutoff.
 }
\end{figure}

Seki and van Kolck~\cite{Seki:2005ns} obtained the values of the
constants,
\begin{equation}
 \theta_1=1.58796\ldots, \quad
  \theta_3=\frac{2}{\pi}, \quad
  R(0) =0.754330\ldots,
\end{equation}
with the three-point difference formula (with $\Lambda=\pi/a$), and
$\theta_5$ is easily evaluated as $\theta_5=12/\pi^3$. (The integral
$I_0$ in~\eqref{watson} can be calculated in a closed form. See
Refs.~\cite{Delves200171, JoyceZucker}.) With these parameters, we see that the
nontrivial fixed point is now located at $(X_\star,
Y_\star)=(-0.76602\ldots, 0.17501\ldots)$. The fixed points and the
flow are depicted in Fig.~\ref{lattice_3pt_flow}.

\comment{the flow is not similar to that in the continuum}

Note that the flow is very different from the one in the continuum,
especially in the strong-coupling phase, i.e., the left-hand part of
the figure. This shows precisely a non-universal feature of the
flow.

\comment{improvement in the discretization}

\begin{figure}[h]
 \includegraphics[width=0.9\linewidth,clip]{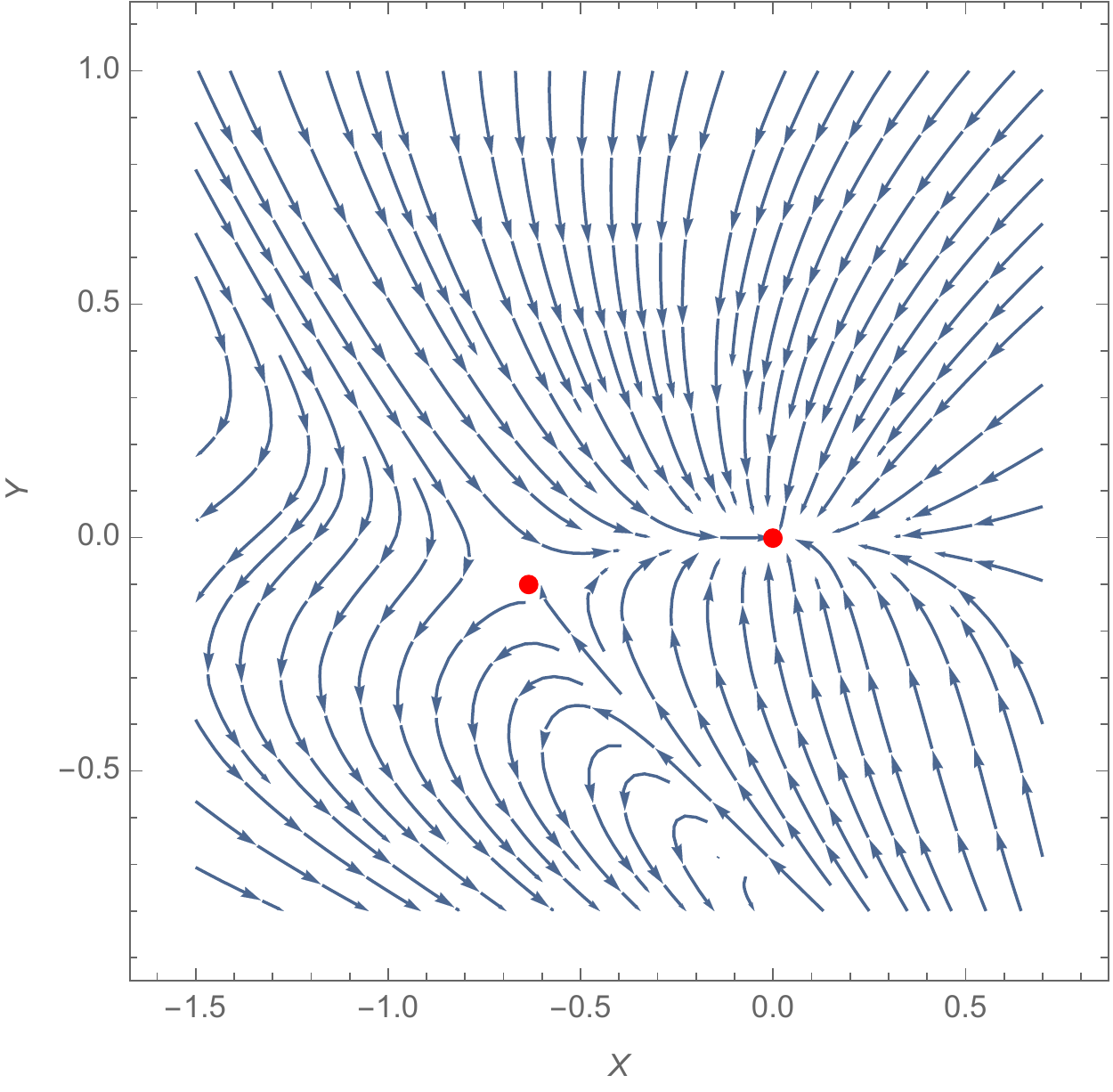}
 \caption{\label{lattice_5pt_flow} The flow and the fixed points of the
 NLO NEFT in the $X$-$Y$ plane obtained by using a lattice
 regularization with the 5-point formula.
 % The allows indicate the directions of
 % the smaller values of the cutoff.
 }
\end{figure}

As we will show in the next section, effects of the rotational
symmetry breaking by the discretization with the three-point formula
are large. We therefore use the five-point difference formula in the
RG analysis. In this case, the values of the constants are
\begin{align}
 \theta_1&=1.37619\ldots, \quad
 \theta_3=\frac{2}{\pi}, \quad
 \theta_5=\frac{15}{\pi^3},\notag \\
 R(0)&=-0.41278\ldots.
\end{align}
Here we have obtained the constants $\theta_1$ and $R(0)$ by a method
similar to that of Appendix of Ref.~\cite{Seki:2005ns}; see
Appendix~\ref{sec:constants} for the detail. In this case, the
nontrivial fixed point is located at $(X_\star,
Y_\star)=(-0.63338\ldots, -0.098805\ldots)$. The fixed points and the flow
are depicted in Fig.~\ref{lattice_5pt_flow}.

\comment{the flow comes closer to the continuum}

The flow changes considerably from the case of the three-point
formula, and gets more similar to the flow in the continuum, as one
might expect.

\comment{deuteron}

\begin{figure}[h]
 \includegraphics[width=0.9\linewidth,clip]{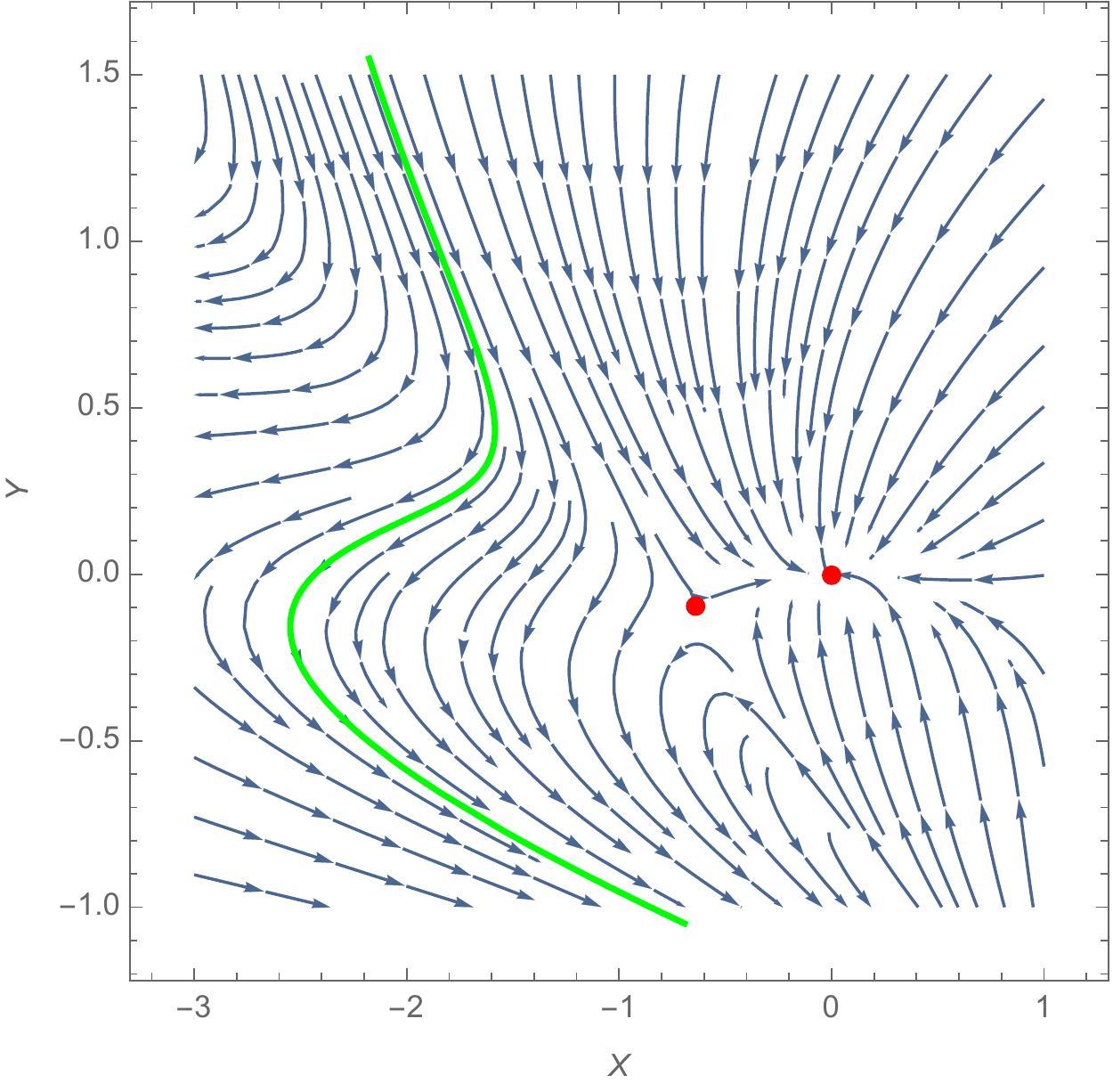}
 \caption{\label{physicalflow_5pt} The flow line corresponding to the
 deuteron obtained by using a lattice
 regularization with the 5-point formula.
 }
\end{figure}

The flow line corresponding to the deuteron is drawn in
Fig.~\ref{physicalflow_5pt}.

\section{Diagonalization of lattice Hamiltonian}
\label{sec:lattice}

\subsection{Lattice Hamiltonian}

\comment{lattice}

In this section, we consider the Hamiltonian diagonalization of the
NLO NEFT without pions on a spatial cubic lattice of a finite lattice
constant $a$ and a finite size $L= N_s a$ with the periodic boundary
condition. The three-dimensional position vector $\bfx$ is replaced by
$\bfn a$, where $\bfn$ is a three-dimensional vector with integer
components $\bfn=(n_1, n_2, n_3)$. The periodic boundary condition
identifies $\bfn$ with $\bfn +N_s \bfe_i$, where $\bfe_i$ ($i=1,2,3$)
is the unit vector in the $i$-th direction.

\comment{Hamiltonian on a lattice}

The Hamiltonian in the continuum,
\begin{align}
 H&=\int d^3x
  \bigg[
   N^\dagger\left(-\frac{\nabla^2}{2M}\right)N
   +C_0 (N^TP_k N)^\dagger (N^T P_k N)
 \notag \\
 &{}\qquad
 -C_2
 \left\{
 (N^TP_k N)^\dagger (N^T P_k\overleftrightarrow{\nabla}^2 N) + \text{h.c.}
 \right\}
 \bigg],
\end{align}
can be transformed into the lattice form $H_L$ by the substitutions,
$\bfx \to \bfn a$, $\int d^3x \to a^3\sum_{\bfn}$, $H\to H_L a^{-1}$,
$N(\bfx)\to N_{\bfn} a^{-3/2}$, $M\to M_L a^{-1}$, $C_0 \to
C_0^{L}a^2$, and $C_2\to C_2^{L}a^4$. The (dimensionless) Hamiltonian
on a lattice is written in terms of dimensionless quantities, and is
given by
\begin{align}
 H_L &= \sum_{\bfn}
  \bigg[
   -\frac{1}{2M_L}N_{\bfn}^\dagger
   %\sum_{i=1}^{3}
 \nabla_L^2 N_{\bfn}
   %\left(N_{\bfn+\bfe_i}-2N_{\bfn}+N_{\bfn-\bfe_i}\right)
   \notag \\
 &{}\qquad\quad + C^{L}_0 
  (N_{\bfn}^TP_k N_{\bfn})^\dagger (N_{\bfn}^T P_k N_{\bfn})
 \notag \\
 &{}\qquad\quad -C_2^{L}
 \left\{
 (N_{\bfn}^TP_k N_{\bfn})^\dagger
 (N_{\bfn}^T P_k\overleftrightarrow{\nabla}_L^2 N_{\bfn}) + \text{h.c.}
 \right\}
 \bigg],
  \label{latH}
\end{align}
where $\nabla_L^2$ represents the discretization of the
(dimensionless) Laplacian. It is given by
\begin{equation}
 \nabla_L^2 N_{\bfn} =\sum_{i=1}^3
  \left(
   N_{\bfn+\bfe_i}-2N_{\bfn}+N_{\bfn-\bfe_i}
  \right)
\end{equation}
for the three-point formula and
\begin{align}
 \nabla_L^2 N_{\bfn}&=\sum_{i=1}^{3}
  \bigg(
   -\frac{1}{12}N_{\bfn+2\bfe_i}+\frac{4}{3}N_{\bfn+\bfe_i}
   -\frac{5}{2}N_{\bfn}
   \notag \\
 &{}\qquad\qquad
 +\frac{4}{3}N_{\bfn-\bfe_i}
   -\frac{1}{12}N_{\bfn-2\bfe_i}
  \bigg)
\end{align}
for the five-point formula.  Similarly, the
$N_{\bfn}^T\overleftrightarrow{\nabla}_L^2 N_{\bfn}$ is given by
\begin{align}
 N_{\bfn}^T\overleftrightarrow{\nabla}_L^2 N_{\bfn}&=
 \sum_{i=1}^{3}\bigg[
 \left(N_{\bfn+\bfe_i}^{T}-2N_{\bfn}^{T}+N_{\bfn-\bfe_i}^{T}\right)P_kN_{\bfn}
 \notag \\
 &{}\quad
 -(N_{\bfn+\bfe_i}^{T}-N_{\bfn}^{T})P_k(N_{\bfn+\bfe_i}-N_{\bfn})
 \notag \\
 &{}\quad
 -(N_{\bfn}^{T}-N_{\bfn-\bfe_i}^{T})P_k(N_{\bfn}-N_{\bfn-\bfe_i})
 \notag \\
 &{}\quad
 +N_{\bfn}^{T}P_k\left(N_{\bfn+\bfe_i}-2N_{\bfn}+N_{\bfn-\bfe_i}\right)
 \bigg]
\end{align}
for the three-point formula and 
\begin{widetext}
 \begin{align}
 N_{\bfn}^T\overleftrightarrow{\nabla}_L^2 N_{\bfn}&=-\frac{1}{12}
 \sum_{i=1}^{3}
  \bigg[
  (N_{\bfn+2\bfe_i}^{T}-16N_{\bfn+\bfe_i}^{T}+30 N_{\bfn}^{T}
  -16N_{\bfn-\bfe_i}^{T} +N_{\bfn-2\bfe_i}^{T})P_k N_{\bfn}
 \notag \\
 &{}\qquad\qquad\qquad
 -\half(N_{\bfn+\bfe_i}^{T}-N_{\bfn}^{T})P_k
 (N_{\bfn+2\bfe_i}-15N_{\bfn+\bfe_i}+15N_{\bfn}-N_{\bfn-\bfe_i})
   \notag \\
 &{}\qquad\qquad\qquad
  -\half(N_{\bfn}^{T}-N_{\bfn-\bfe_i}^{T})P_k
 (N_{\bfn+\bfe_i}-15N_{\bfn}+15N_{\bfn-\bfe_i}-N_{\bfn-2\bfe_i})
  \notag \\
 &{}\qquad\qquad\qquad
  -\half(N_{\bfn+2\bfe_i}^{T}-15N_{\bfn+\bfe_i}^{T}+15N_{\bfn}^{T}
  -N_{\bfn-\bfe_i}^{T})P_k(N_{\bfn+\bfe_i}-N_{\bfn})
  \notag \\
 &{}\qquad\qquad\qquad
  -\half(N_{\bfn+\bfe_i}^{T}-15N_{\bfn}^{T}+15N_{\bfn-\bfe_i}^{T}
  -N_{\bfn-2\bfe_i}^{T})P_k(N_{\bfn}-N_{\bfn-\bfe_i})
  \notag \\
 &{}\qquad\qquad\qquad
  +N_{\bfn}^{T}P_k(N_{\bfn+2\bfe_i}-16N_{\bfn+\bfe_i}+30 N_{\bfn}
  -16 N_{\bfn-\bfe_i}+N_{\bfn-2\bfe_i})
  \bigg]
\end{align}
\end{widetext}
for the five-point formula.

\comment{momentum space}

It is easier to work in momentum space. We then Fourier transform the
nucleon operator as
\begin{equation}
 N_{\bfn} = \frac{1}{N_{s}^{3/2}}\sum_{\bfp} e^{i\bfp\cdot\bfn} a_{\bfp},
  \label{fourierN}
\end{equation}
where we suppress the spin and isospin indices.
The momentum $\bfp=(p_1,p_2,p_3)$ takes the values
\begin{equation}
 p_i = \frac{2\pi}{N_s}\hat{p}_i,
\end{equation}
where integers $\hat{p}_i\ (i=1,2,3)$ satisfy
\begin{equation}
 -\frac{N_s}{2} < \hat{p}_i \le \frac{N_s}{2}.
\end{equation}
The creation and annihilation operators satisfy the canonical
anti-commutation relations,
\begin{equation}
 \{a_{\bfp}, a_{\bfp'}\}=\{a^\dagger_{\bfp},a^\dagger_{\bfp'}\}=0,
  \quad
 \{a_{\bfp}, a^\dagger_{\bfp'}\} =\delta_{\hat{\bfp},\hat{\bfp}'}.
\end{equation}

\comment{hamiltonian in terms of creation and annihilation operators}

By substituting Eq.~\eqref{fourierN} into Eq.~\eqref{latH}, we obtain
the Hamiltonian in terms of creation and annihilation operators,
\begin{align}
 H_L &=\sum_{\bfp} \frac{\Delta_{\bfp}}{2M_L} a_{\bfp}^{\dagger} a_{\bfp}
 %\notag \\
 %&
 +\frac{1}{N_s^3} \sum_{\left\{\bfp_i\right\}}
  \delta_{\bfp_1+\bfp_2-\bfp_3-\bfp_4,\bm{0}}
 \notag \\
 &{}\quad
 \times
 \bigg[
 C_{0}^{L}+C_2^{L}\left(\Delta_{\bfp_1,\bfp_2}+\Delta_{\bfp_3,\bfp_4}\right)
 \bigg]
  \notag \\
 &{}\quad
 \times(a^\dagger_{\bfp_1}P_k a^\dagger_{\bfp_2})(a_{\bfp_3}P_k a_{\bfp_4}),
\end{align}
where
\begin{equation}
 \Delta_{\bfp} =
  \begin{cases}
   \displaystyle
   4\sum_{i=1}^{3}\sin^2\left(\frac{p_i}{2}\right), \\
   \qquad\qquad\qquad\qquad
   (\text{three-point formula})\\
   \displaystyle
   4\sum_{i=1}^{3}
   \left[
   \sin^2\left(\frac{p_i}{2}\right)
   +\frac{1}{3}\sin^4\left(\frac{p_i}{2}\right)
   \right], \\
   \qquad\qquad\qquad\qquad
   (\text{five-point formula})
  \end{cases}
\end{equation}
and
\begin{equation}
 \Delta_{\bfp,\bfq}=
  \begin{cases}
   \displaystyle
   4\sum_{i=1}^{3}
   \bigg(
   \sin^2\left(\frac{p_i}{2}\right)
   +\sin^2\left(\frac{q_i}{2}\right) \\
   \qquad
   \displaystyle
   -2\cos\left(\frac{p_i+q_i}{2}\right)
   \sin\left(\frac{p_i}{2}\right)
   \sin\left(\frac{q_i}{2}\right)
   \bigg), \\
   \qquad\qquad\qquad\qquad
   (\text{three-point formula})\\
   \displaystyle
   4\sum_{i=1}^{3}
   \bigg(
   \sin^2\left(\frac{p_i}{2}\right)
   +\frac{1}{3}\sin^4\left(\frac{p_i}{2}\right)\\
   \qquad\ 
   \displaystyle
   +\sin^2\left(\frac{q_i}{2}\right)
   +\frac{1}{3}\sin^4\left(\frac{q_i}{2}\right)\\
   % crossterms
   \qquad\ 
   \displaystyle
   -\cos\left(\frac{p_i+q_i}{2}\right)\\
   \qquad
   \displaystyle
   \times
   \bigg\{
   \sin\left(\frac{p_i}{2}\right)
   \left(
   \sin\left(\frac{q_i}{2}\right)
   +\frac{1}{3}\sin^3\left(\frac{q_i}{2}\right)
   \right)
    \\
   \qquad\ 
   \displaystyle
   +
   \left(
   \sin\left(\frac{p_i}{2}\right)
   +\frac{1}{3}\sin^3\left(\frac{p_i}{2}\right)
   \right)\sin\left(\frac{q_i}{2}\right)
   \bigg\}
   \bigg).\\
   \qquad\qquad\qquad\qquad
   (\text{five-point formula})
  \end{cases}
\end{equation}
Note that $\Delta_{\bfp,\bfq}=\Delta_{\bfq,\bfp}$ and
$\Delta_{\bfp,-\bfp}=4\Delta_{\bfp}$.

\subsection{Schr\"odinger equation for two-nucleon states}

\comment{Schr\"odinger eq. for a two-nucleon state}

Now we consider the lattice version of the (stationary) Schr\"odinger
equation,
% \begin{equation}
%  H|\Psi\ket =E|\Psi\ket.
% \end{equation}
%  On a lattice, the Schr\"odinger equation takes the form
\begin{equation}
 H_L |\Psi^k\ket = E_L |\Psi^k\ket,
\end{equation}
where $|\Psi^k\ket$ is the two-nucleon state with the zero total
momentum and the spin-triplet isospin-singlet
%$^3S_1$
projection,
\begin{equation}
 |\Psi^k\ket =\sum_{\bfp}\psi(\bfp)
  a_{\bfp}^\dagger P_k^{\dagger} a_{-\bfp}^\dagger |0\ket,
\end{equation}
and $E_L = E a$ is the dimensionless energy eigenvalue. In terms of
the discretized ``momentum-space wavefunction'' $\psi(\bfp)$ of
relative motion, the Schr\"odinger equation can be written as
\begin{align}
 E_L\psi(\bfp) &=\frac{\Delta_{\bfp}}{M_L} \psi(\bfp)
 \notag \\
 &{}\quad
 +\frac{1}{N_s^3}\sum_{\bfq}
  \left[
 C_0^{L}+4C_2^{L}\left(\Delta_{\bfp}+\Delta_{\bfq}\right)
 \right]\psi(\bfq),
  \label{EigenEq}
\end{align}
which is nothing but the discretized version of
Eq.~\eqref{continuumSchroedinger}.

\comment{diagonalization}

We numerically diagonalize the eigenvalue
equation~\eqref{EigenEq}. The physical length of the lattice constant
is determined by giving $M_L$ through the relation $M_L = M a$, where
$M$ is the physical nucleon mass which we set $M=938.9$ MeV. (Note
that there is no self-energy contribution in our theory.) We
typically consider the case $a=5$ fm, which corresponds to the
momentum cutoff $\Lambda=\pi/a\approx 124$ MeV. Most of the
calculations are done with $N_s=16$, which corresponds to a cube
with the edge of length $L=80$ fm.

\comment{interested only in groundstate}

We are interested only in the groundstate. In the strong coupling
phase, it is a boundstate. In the weak coupling phase, there is no
boundstate physically, but the periodic boundary condition makes the
groundstate have negative energy~\cite{Luscher:1986pf}.

\subsection{RG analysis}
\label{sec:lattRG}

In the following analysis, we use the dimensionless coupling constants
$X$ and $Y$ defined in Eq.~\eqref{c0c2}, but with $\Lambda=\pi/a$.

\comment{low energy quantities}

We choose the binding energy and the ANC as low energy physical
quantities and require them to be independent of the lattice
constant. The RG flow can be numerically obtained by first calculating
the binding energy and the ANC for a set of $(X,Y)$ and then changing
the lattice constant a bit from $a$ to $a+\delta a$ and searching
numerically the new set of $(X+\delta X,Y+\delta Y)$ that gives the
same binding energy and the ANC.

\comment{how to calculate the ANC}

The ANC is most easily obtained by fitting the numerically obtained
(normalized) ``momentum-space wavefunction'' $\psi(\bfp)$ to the
expression
\begin{equation}
 \psi(\bfp)=A+\frac{B}{M_L|E_L|+\Delta_{\bfp}},
  \label{defANC}
\end{equation}
and determining the constants $A$ and $B$. Note that this form of the
wavefunction is implied by the Schr\"odinger equation~\eqref{EigenEq},
and corresponds to the continuum wavefunction,
Eq.~\eqref{continuumwf}. We thus identify $B/4\pi$ with the ANC.

\comment{the flow}

\begin{figure}[h]
 \includegraphics[width=0.9\linewidth,clip]{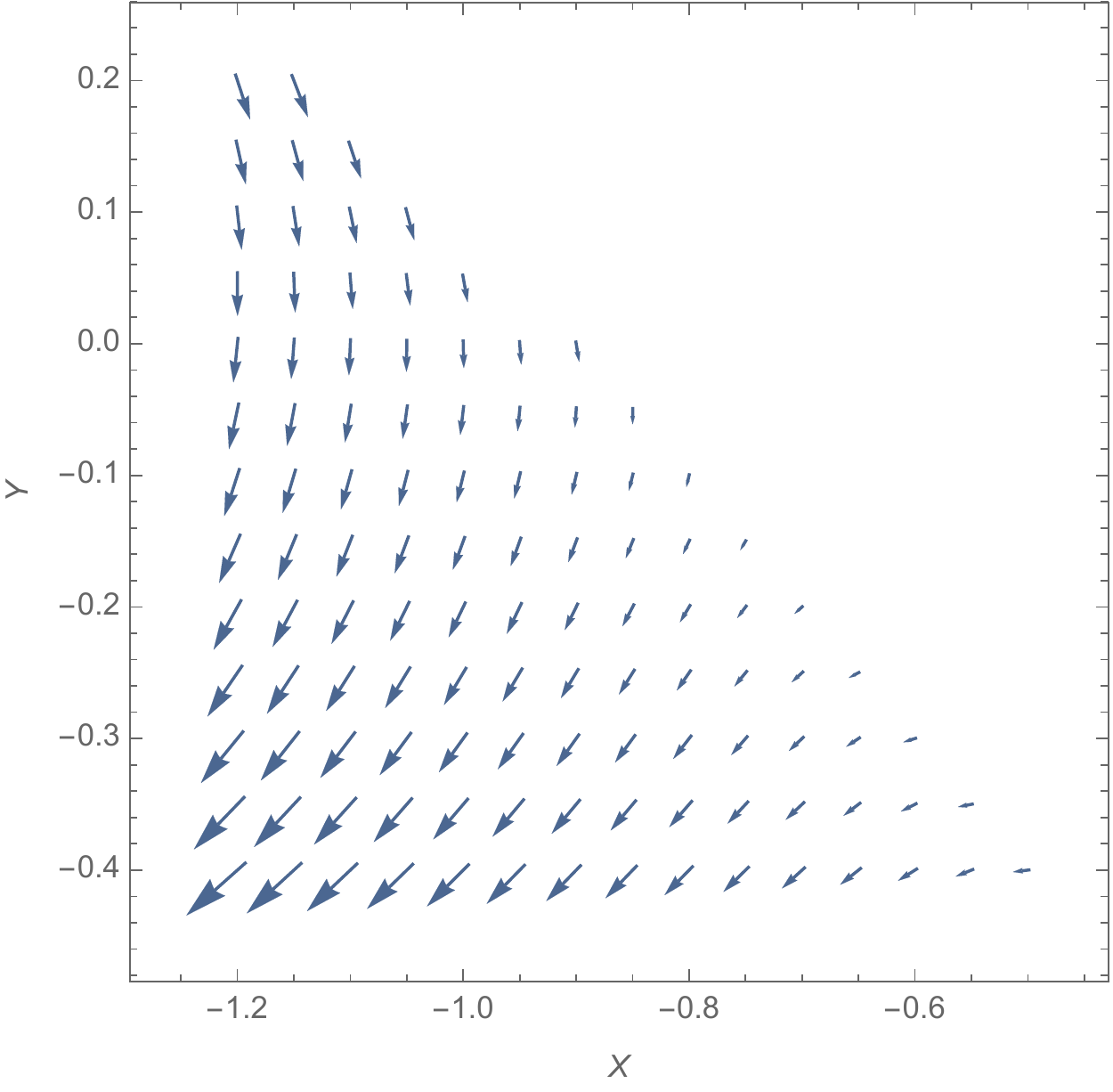}
 \caption{\label{lattice_genuine_flow} The flow of the NLO NEFT in the
 strong coupling phase in the $X$-$Y$ plane obtained by numerical
 diagonalization of the Hamiltonian defined on a lattice with the
 5-point formula.
 % The allows indicate the directions of
 % the smaller values of the cutoff.
 }
\end{figure}

In Fig.~\ref{lattice_genuine_flow}, we show the RG flow calculated with
the five-point formula. We draw the change $(\delta X, \delta Y)$ for
$a=5$ fm and $\delta a=0.25$ fm.

\comment{we do not consider the weak coupling phase}

The right upper part of the figure corresponds to the weak coupling
phase. Because of the fictitious feature of the groundstate energy in
the weak coupling phase, we do not calculate the flow.

\comment{it is difficult to calculate the flow near the phase
boundary}

It is difficult to calculate the flow near the phase boundary. Near
the phase boundary, the groundstate energy becomes small, and the
effects of the periodic boundary condition becomes
noticeable~\cite{Luscher:1985dn}. The wavefunction with the radius
$\sim L/2= 40$ fm is affected by the boundary condition. This radius
corresponds to the binding energy 0.03 MeV. The finite-size
effect however brings about useful information, as shown below.

\comment{L dependence of the ground state energy}

\begin{figure}[h]
 \includegraphics[width=0.9\linewidth,clip]{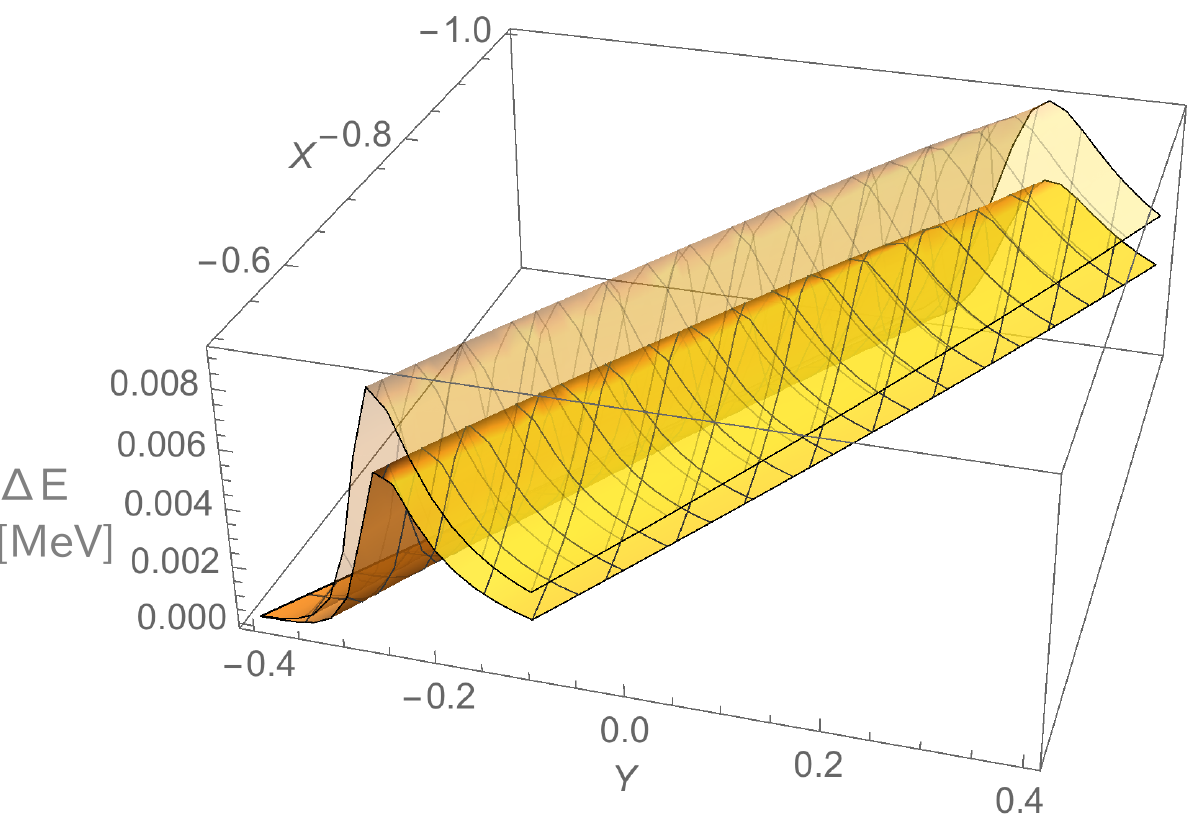}
 \caption{\label{L-dep} The difference of the calculated groundstate
 energies with $N_s=14$ and $N_s=16$ (upper surface), and that with
 $N_s=16$ and $N_s=18$ (lower surface) are shown as functions of $X$
 and $Y$. This side of the mountain range is the weak coupling phase,
 the other side the strong coupling phase.}
\end{figure}

The $L$ dependence of the groundstate is shown in Fig.~\ref{L-dep},
where the difference of the calculated groundstate energies with $N_s=14$ and
$N_s=16$, and the difference with $N_s=16$ and $N_s=18$ are
plotted. Note that the difference is larger in the $Ns=14$
v.s. $N_s=16$ case than in the $Ns=16$ v.s. $N_s=18$ case, as one
naturally expects.

\comment{physical meaning of the ridge line}

\begin{figure}[h]
 \includegraphics[width=0.9\linewidth,clip]{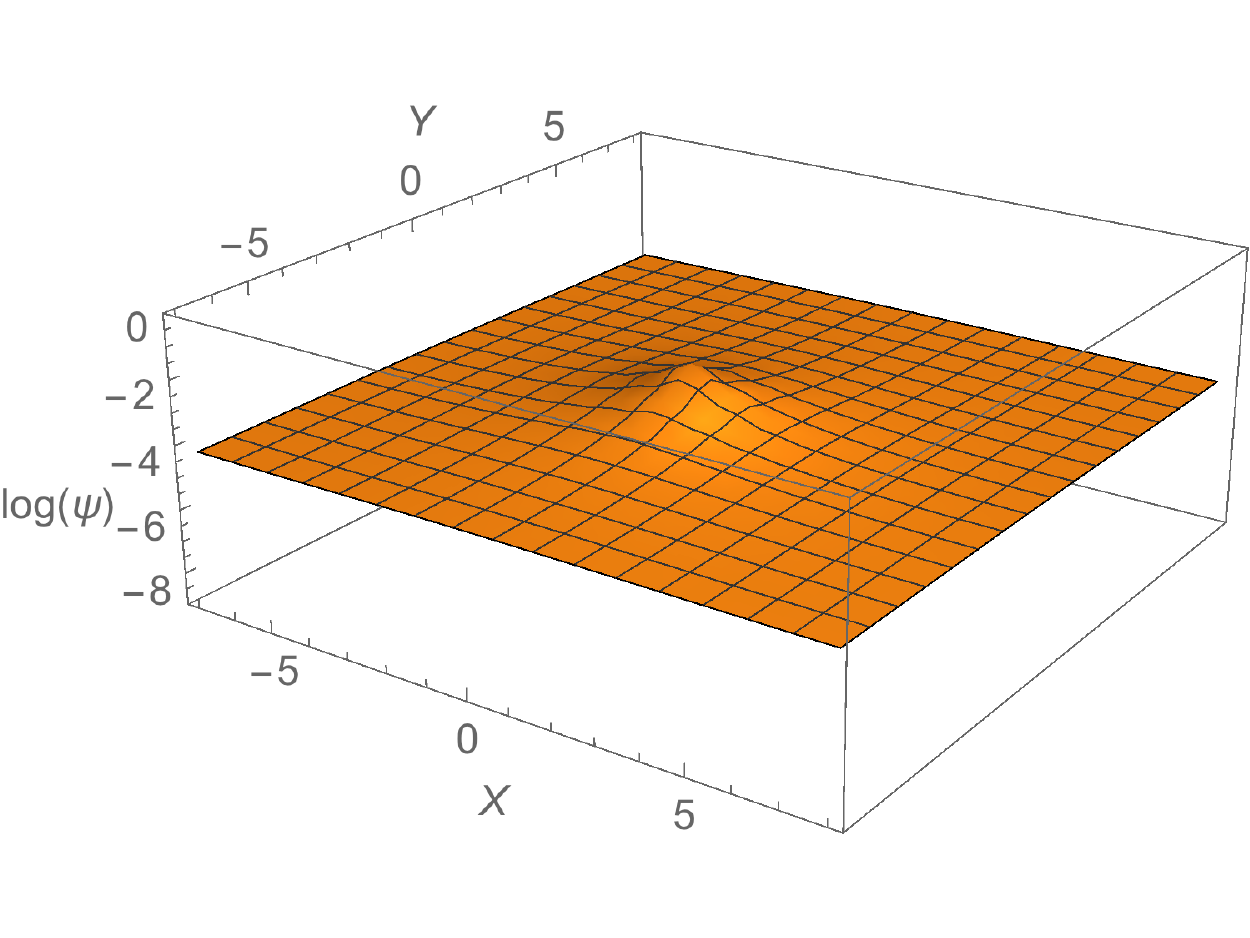}
 \includegraphics[width=0.9\linewidth,clip]{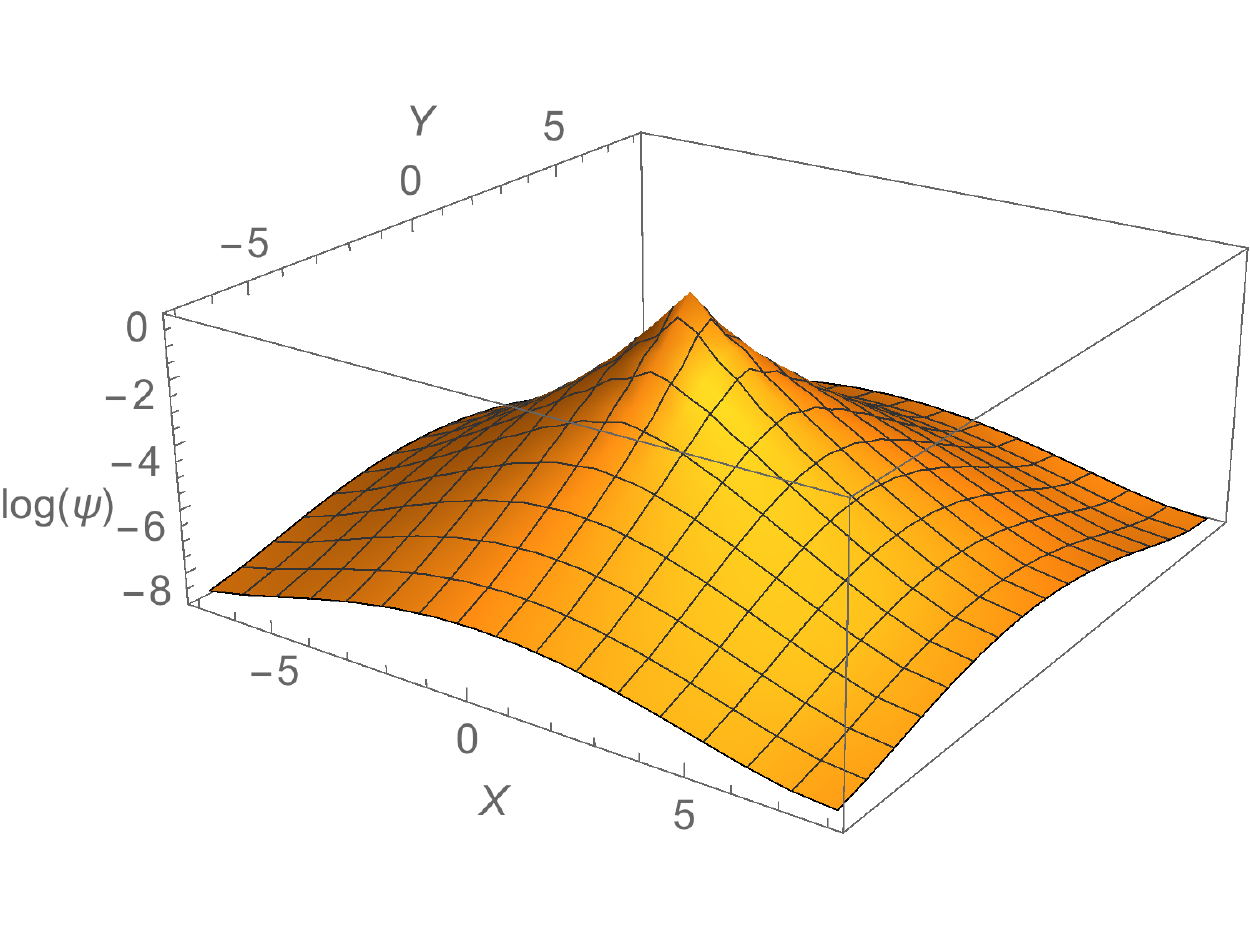}
 \caption{\label{wavefunctions} Typical wavefunctions near the
 critical line. In the weak coupling phase (upper), the wavefunction
 spreads out over the whole space with a small peak at the center of
 potential. In the strong coupling phase (lower), it is sharply peaked
 at the center of the potential and decays exponentially.}
\end{figure}

It is numerically shown that the ridge line is $L$ independent. We
argue that this ridge line represents the phase boundary. First note
that, as we discussed above, the $L$ dependence of the groundstate
energy in the strong coupling phase comes from the spreading of the
wavefunction as we approach the critical line (phase boundary). The
energy difference becomes therefore larger as we approach the critical
line. Second, in the weak coupling phase, the $L$ dependence arises
for a different reason; the wavefunction in the weak coupling phase
spreads out over the whole space and feels periodically placed
potentials. The smaller the period is, the more negative the
groundstate energy is, because the ``density'' of the attractive
potential is higher when the period is smaller. The $L$ dependence of
the groundstate energy in the weak coupling phase is weaker than that
in the strong coupling phase. We show the typical wavefunctions in the
strong and  weak coupling phases in
Fig.~\ref{wavefunctions}. Finally, $L$ dependence of calculated
groundstate energies fit well with the known $L$ dependence
of Refs.~\cite{Luscher:1985dn, Luscher:1986pf} for the both sides
of the ridge line.

\comment{combined}

\begin{figure}[h]
 \includegraphics[width=0.9\linewidth,clip]{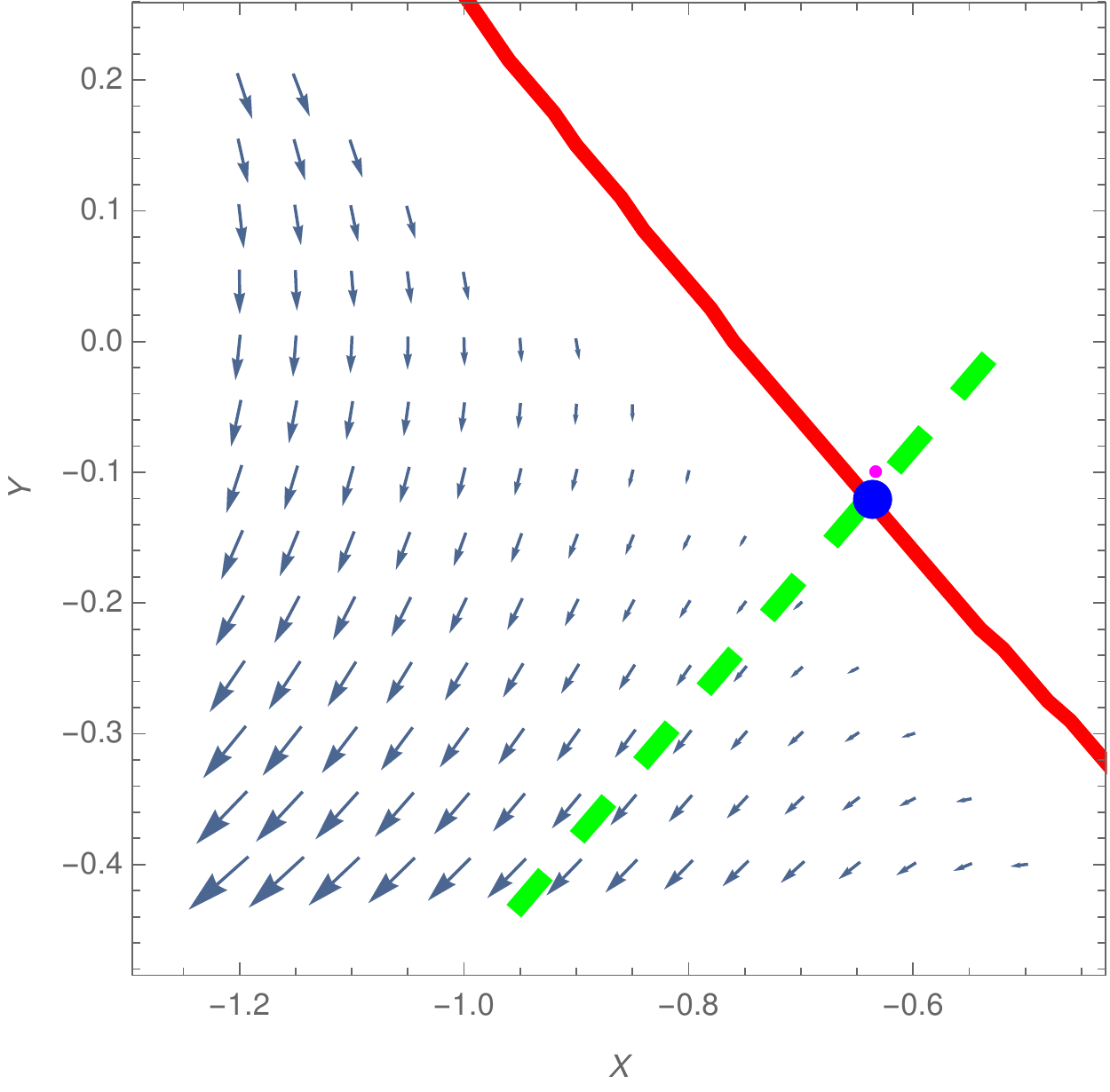}
 \caption{\label{combined5}The ridge line (red) together with the RG
 flow in Fig.~\ref{lattice_genuine_flow} calcuated with the five-point
 formula. From the flow, we infer that the nontrivial fixed point is
 on the dashed line (green). The nontrivial fixed point is also on the
 ridge line, it is at the crossing point (blue bullet). The small
 point (magenta) just above the crossing point is the location of the
 nontrivial fixed point obtained by the analytic calculation,
 $(X_\star, Y_\star)=(-0.63338\ldots, -0.098805\ldots)$.}
\end{figure}

Once we establish that the ridge line represents the phase boundary,
it is easy to locate the nontrivial fixed point. In
Fig.~\ref{combined5}, we show the ridge line together with the RG
flow. The RG flow indicates the \textit{direction} in which the
nontrivial fixed point resides. In addition, it is on the phase
boundary. These allow us to identify where the nontrivial fixed point
is. It is $(X_\star,Y_\star)=(-0.65\sim -0.63, -0.13\sim -0.11)$,
which is surprisingly close to the one obtained analytically with the
five-point formula in Sec.~\ref{sec:analytic}.

\comment{''direction'' represents the relevant operator}

The direction in which the RG flow goes out from the nontrivial fixed
point (the dashed line direction in Fig.~\ref{combined5}) represents
the relevant operator. The (unit) vector for the direction is
$(-1/\sqrt{2},-1/\sqrt{2})$ within the accuracy of the present
analysis. This is very different from $(-0.933\ldots, -0.359\ldots)$
obtained by linearizing the RGEs~\eqref{rgeX} and \eqref{rgeY} with
the five-point formula around the nontrivial fixed point.

\comment{for the 3-point formula}

\begin{figure}[h]
 \includegraphics[width=0.9\linewidth,clip]{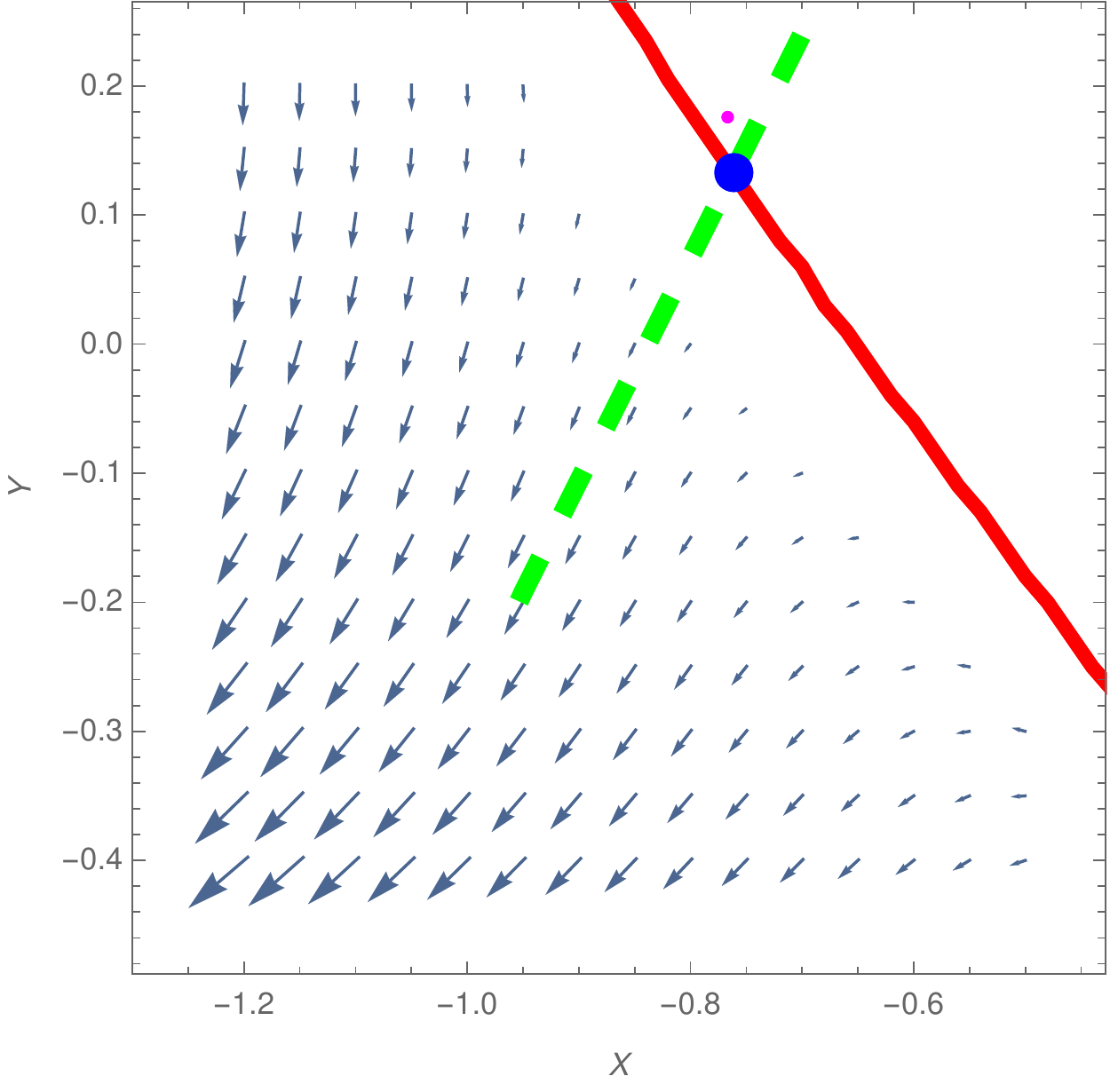}
 \caption{\label{combined3} The same as in Fig.~\ref{combined5}, but
 with the three-point formula. The small point (magenta) indicates the
 location of the nontrivial fixed point obtained by the analytic
 calculation, $(X_\star,Y_\star) =(-0.76602\ldots, 0.17501\ldots)$.}
\end{figure}

We perform similar analysis with the three-point formula. In
Fig.~\ref{combined3} we show the ridge line together with the RG flow.
The RG flow is considerably different from that with the five-point
formula. The location of the nontrivial fixed point is
$(X_\star,Y_\star)=(-0.75 \sim -0.77, 0.12 \sim 0.14)$. It is again
very close to the one analytically obtained. The relevant direction is
now represented by a vector $(-1/2,-\sqrt{3}/2)$ within the accuracy
of the present analysis. It should be compared with $(-0.935\ldots
,0.353\ldots)$ obtained from the linearized RGEs derived from
Eqs.~\eqref{rgeX} and \eqref{rgeY} with the three-point formula.

\comment{to summarize}

To summarize, the analytic results with the lattice regularization,
which are not obtained on a lattice, are very accurate for the
location of the nontrivial fixed point, but the direction of the
relevant operator is considerably different from the one on a lattice.

\iftrue
\comment{explicit rotational symmetry breaking}

It is instructive to see how the explicit rotational-symmetry breaking
affects the shape of the ``wavefunction.'' In Figs.~\ref{anc}, we show
$\psi(\bfr)/(e^{-\sqrt{M_L|E_L|}r}/r)$ in the $(1,0,0)$, $(1,1,0)$,
and $(1,1,1)$ directions as a function of $r=|\bfr|$, where
$\psi(\bfr)$ is the Fourier transform of $\psi(\bfp)$. Precisely, for
the function taken as vertical axis, we have taken into account the
periodicity minimally, that is, the effect of the potentials within
the distance $L=N_s a$, while the potentials at larger distances
give negligibly small corrections and are thus ignored. For
example, the function we actually have plotted in the $(1,0,0)$
direction is $\psi(na,0,0)/(e^{-\sqrt{M_L|E_L|}na}/na\;
+\; e^{-\sqrt{M_L|E_L|}(N_sa-na)}/(N_sa-na))$ for integers $n$
satisfying $0\le n\le N_s$.  If the
wavefunction were rotationally symmetric, they would coincide with
each other and show a plateau (with the value of ANC) at long
distances. We see that the calculation with the three-point formula
shows rather large direction-dependence, but the use of the five-point
formula reduces it largely.

\begin{figure}[h]
 \includegraphics[width=0.9\linewidth,clip]{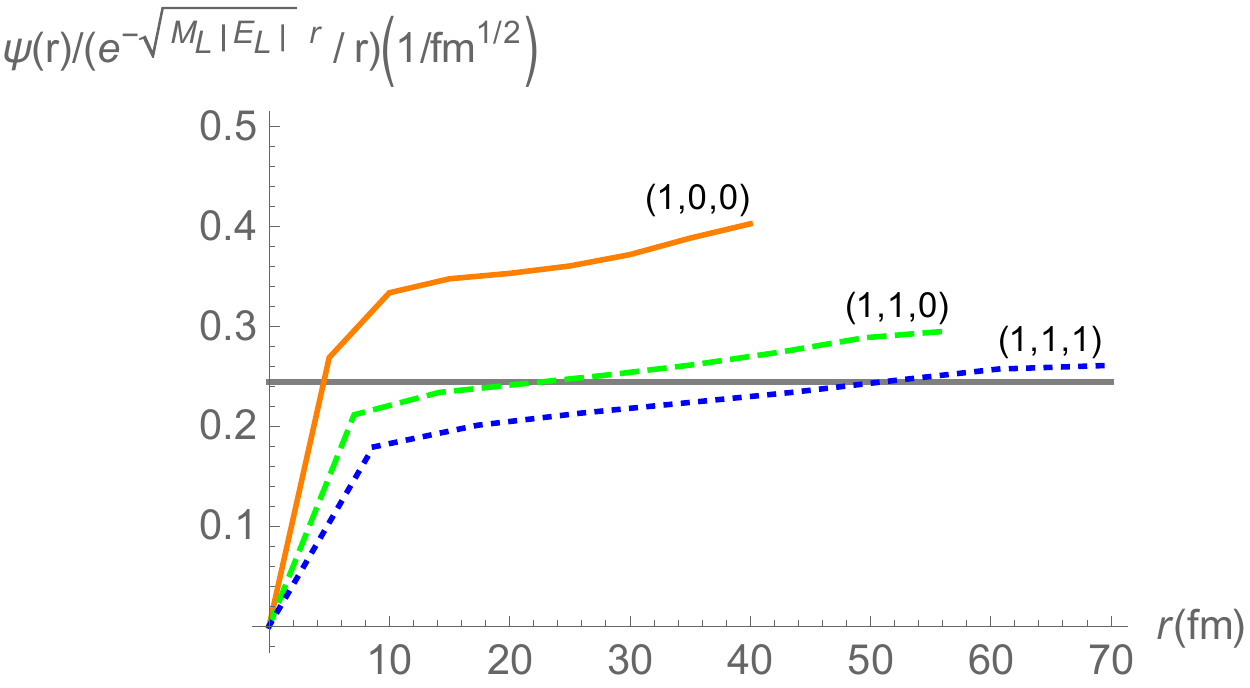}
 \includegraphics[width=0.9\linewidth,clip]{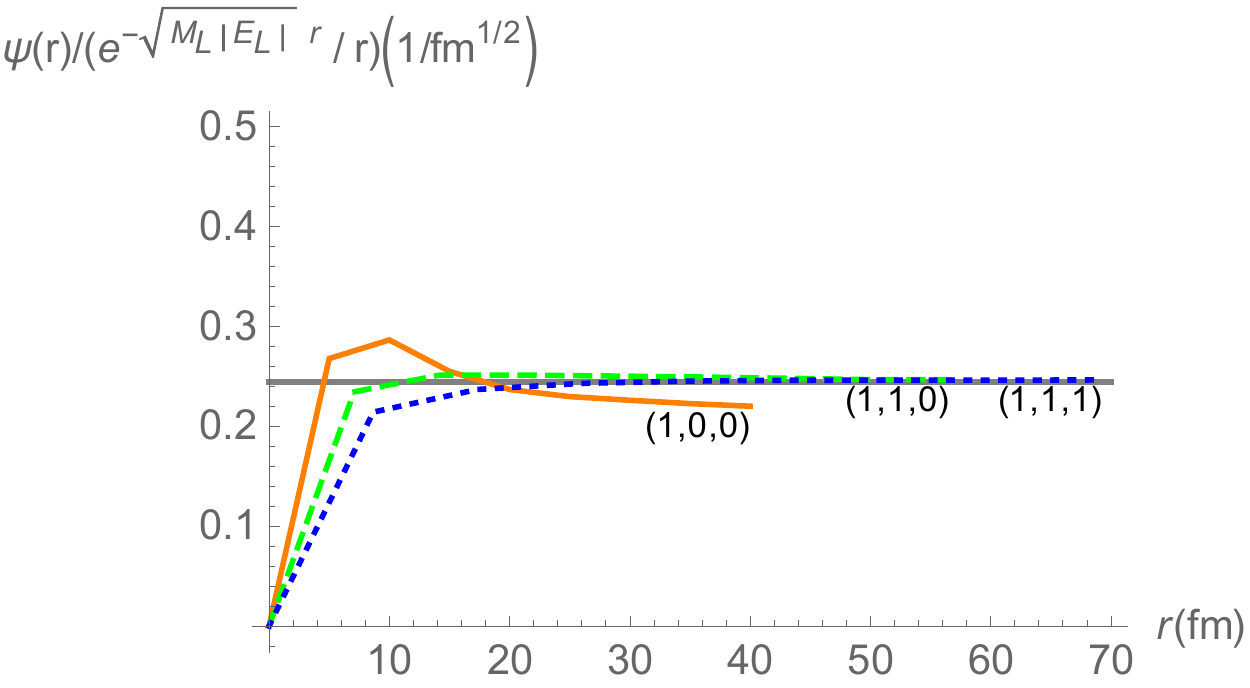}
 \caption{\label{anc} The rotational symmetry breaking in the
 asymptotic behavior of the ``wavefunction.'' The ``wavefunction'' in
 the $(1,0,0)$, $(1,1,0)$, and $(1,1,1)$ directions are obtained by
 the diagonalization of the Hamiltonian with the three-point formula
 (upper) and with the five-point formula (lower). The grey line
 indicates the ANC defined as $B/4\pi$ from the coefficient $B$ of the
 regularized Yukawa term in Eq.~\eqref{defANC}. The calculation is done
 for the deuteron state, so that ANC is 0.224 fm$^{-1/2}$.}.
\end{figure}

\fi

\subsection{The flow line correspoinding to deuteron}

\comment{physical flow line}

\begin{figure}[h]
 \includegraphics[width=0.9\linewidth,clip]{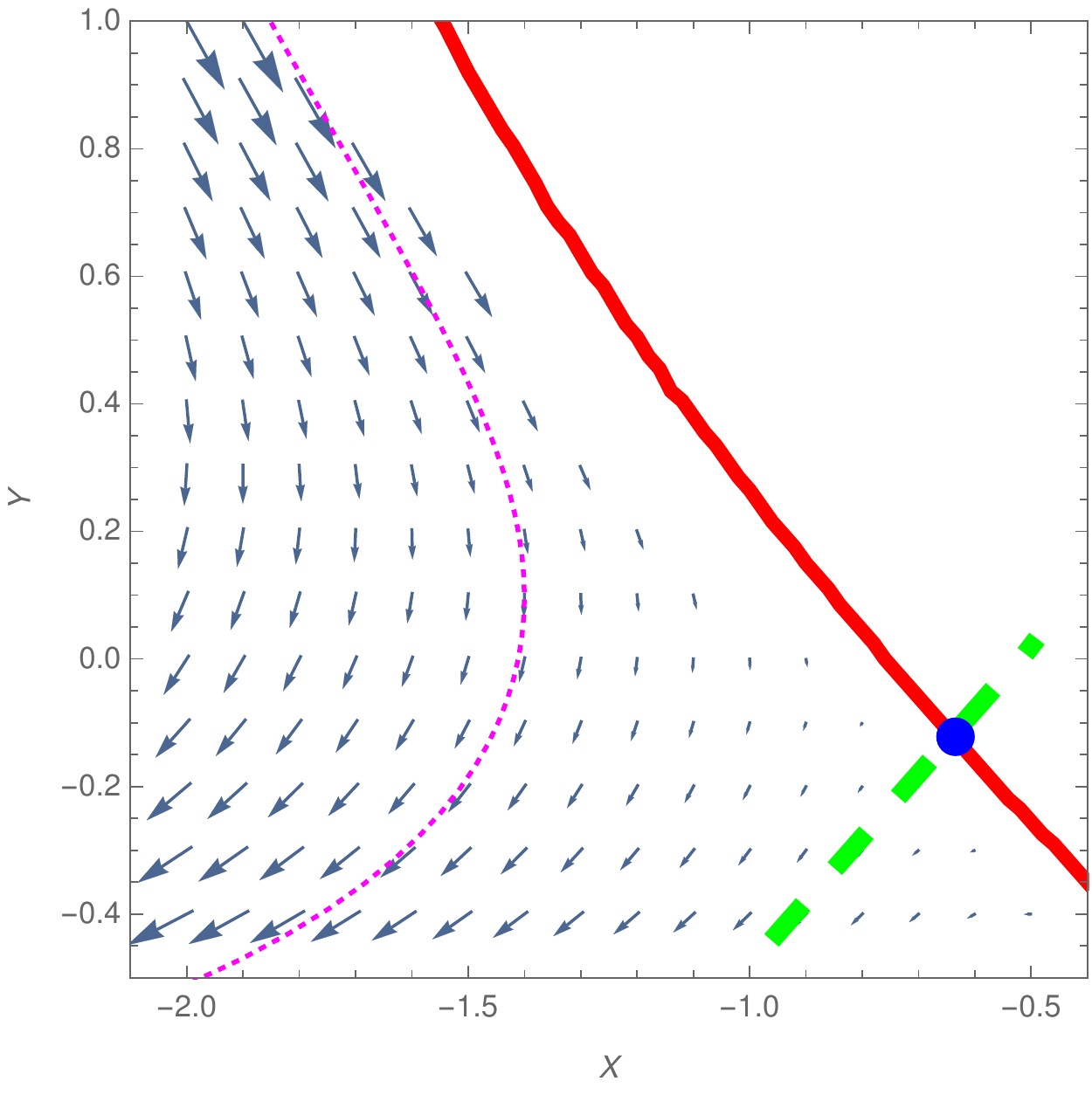}
 \caption{\label{physflow5}The flow line corresponding to deuteron is
 shown as a dotted line (magenta) against the RG flow given in
 Fig.~\ref{combined5}. The calculations are done with the five-point
 formula. The nontrivial fixed point (blue bullet) phase boundary (red
 solid line), and the relevant direction (green dashed line) are also
 shown.}
\end{figure}

\begin{figure}[h]
 \includegraphics[width=0.9\linewidth,clip]{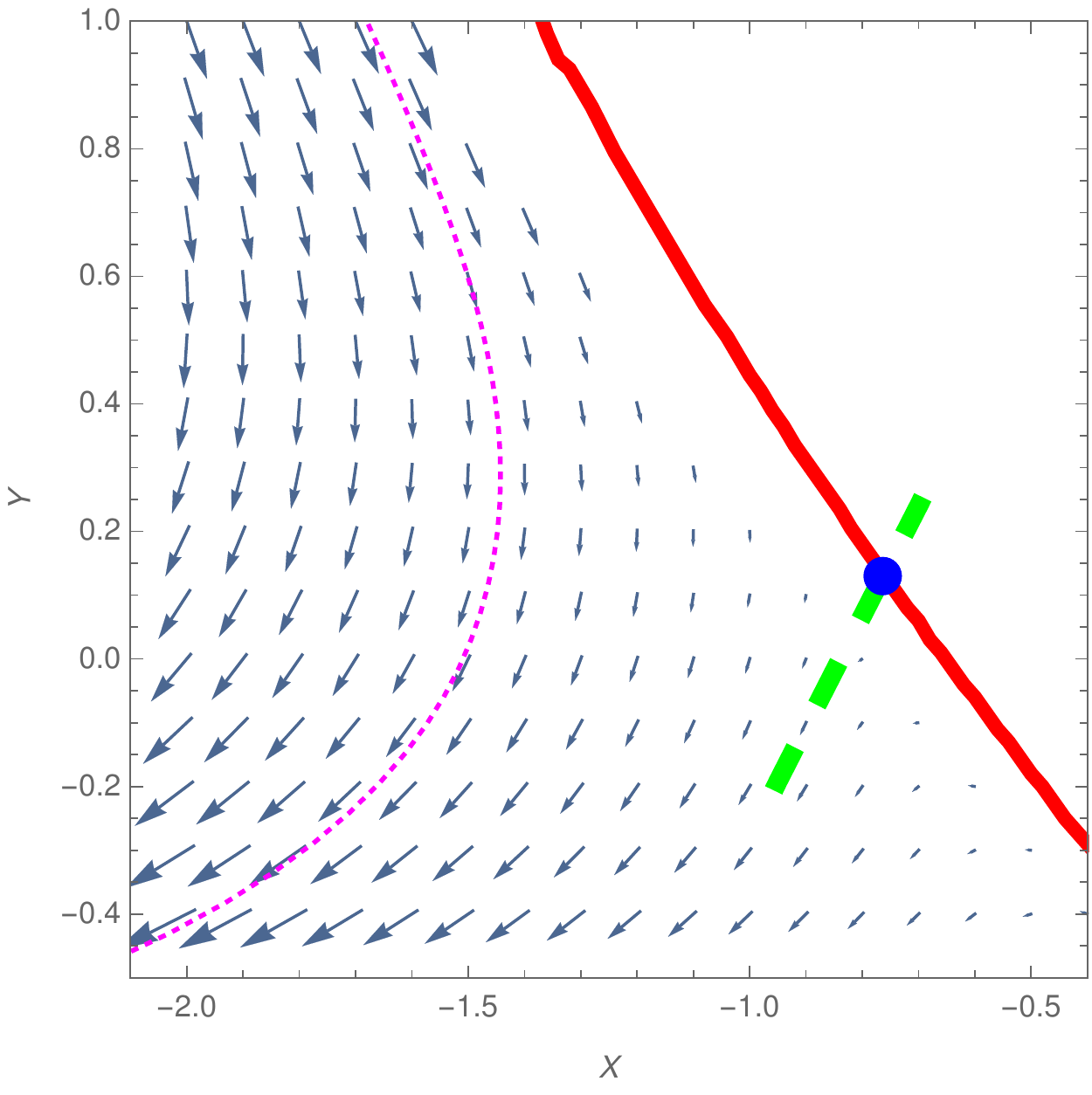}
 \caption{\label{physflow3}The same as in Fig.~\ref{physflow5}, but
 the calculations are done with the three-point formula.}
\end{figure}

Finally we draw a flow line that corresponds to deuteron. As input
parameters, we use the binding energy $E=2.19$ MeV and the ANC $=0.244$
fm$^{-1/2}$, which are obtained in Sec.~\ref{sec:setup}. The flow line
is shown in Fig.~\ref{physflow5} for the five-point formula, and in
Fig.~\ref{physflow3} for the three-point formula, together with the RG
flow, the nontrivial fixed point, the phase boundary, and the relevant
direction.

\comment{how to get it}

The flow line is obtained as follows: (1) We look for a point $(X_0,
Y_0)$ for which the binding energy and the ANC are takes the values
given above for a certain value of the lattice constant $a_0$. (2) We
then change the lattice constant 5\% larger, $a_1=1.05 a_0$, and
search for a new set of coupling constants $(X_1,Y_1)$ for which the
binding energy and the ANC take the same values. (3) We repeat the
procedure; that is, starting with the set of coupling constants $(X_1,
Y_1)$ and the lattice constant $a_1$, we change the lattice constant
5\% larger, $a_2=1.05a_1$, and search for a new set of the coupling
constant $(X_2, Y_2)$ for which the binding energy and the ANC take
the same values, and so on. Remember that our lattice Hamiltonian does
not contain the lattice constant and its value is determined through
the dimensionless nucleon mass $M_L =M a$, so that changing the value
of $M_L$ amounts to changing the value of $a$. When drawing
Figs.~\ref{physflow5} and \ref{physflow3}, we change $M_L$ in the
region $9\alt M_L \alt 80$, corresponding 2 fm $\alt a \alt 17$
fm. Of course the lattice with $a\sim$ 2 fm is too fine for the
present EFT, the calculation there should not take too seriously.

\comment{the part closest to the nontrivial fixed point}

The part of the flow closest to the nontrivial fixed point corresponds
to the lattice constant $a$ in the range $5\sim 10$ fm, corresponding
to the momentum scale $62 \sim 124$ MeV, just in the
region of validty of the EFT without pions.

\section{Summary and discussions}
\label{sec:summary}

\comment{what we did}

In this paper, we diagonalize the Hamiltonian for the NLO NEFT without
pions defined on a spatial lattice in order to obtain the two-nucleon
boundstate, which is mainly in the S wave. We obtain the RG flows by
changing the lattice constant, with the binding energy and the ANC
fixed. By examining the flows, we can infer the relevant operator,
which corresponds to the flow going out from the nontrivial fixed
point. Thus, we know in which direction the fixed point resides. In
addition, we identify where the finite-size effect on the binding
energy is maximal and argue that the line is the phase boundary. The
nontrivial fixed point is known to be on the phase boundary. From
these, we can determine the location of the nontrivial fixed point
numerically. It turns out that it is very close to the point obtained
by the corresponding analytic calculation, with the divergent
integrals in the continuum RG equations being lattice regularized. In
contrast, the relevant operator is considerably different from the
correspoinding one analytically obtained.

\comment{rotational symmetry breaking}

The most of the difference between the analytic results with lattice
regularization and the genuine lattice results may be considered as
the rotational symmetry breaking effects. We show that improving the
representation of derivatives, from the three-point formula to the
five-point formula, tends to reduce the difference.

\comment{ANC}

We also show that the ANC, together with the binding energy, can be
used as a low-energy physical quantity to fix the effective field
theory couplings for a wide range of the cutoff, at least to
investigate the physical system (deuteron).

\comment{ANC is not good for deep boundstates}

The ANC however does not seem to work for deep boundstates. This is
because of a finite cutoff imposed on the EFT, not specific to the
lattice regularization. Remember that the wavefunction may be written
as a sum of a regularized delta function and a regularized Yukawa
function, and our definition of the ANC is the coefficient of the
latter. The regularized Yukawa function damps exponentially and the
damping depends on the binding energy; it damps more rapidly for
larger binding energies. On the other hand, the regularized delta
function damps independently of the binding energy. Thus, for deep
boundstates, the asymptotic form of the wavefunction is dominated by
the regularized delta function. Fixing the ANC there does not control
physics any more. We think that this is the reason why the RG flows
obtained numerically differ considerably from those obtained
analytically in the region of deep boundstates.

\comment{weak coupling phase}

Throughout these analyses, we confine ourselves to the strong coupling
phase and look at the properties of the boundstates.  Calculations in
the weak coupling phase, on the other hand, will bring about
information on physical quantities of scattering through so-called
Luscher's formula~\cite{Luscher:1990ux}. It can be used as inputs for
the coupling constants of the operators in other channels.

\begin{acknowledgments}
 This work was supported by JSPS KAKENHI Grant Numbers 15K05082(K.H.),
 26$\cdot$5861(S.S.), and 26400278(M.Y.).
\end{acknowledgments}

\appendix*

\section{Evaluation of $I_0$ with the five-point formula}
\label{sec:constants}

\comment{the target}

In this section, we consider the integral
\begin{equation}
 W(z)=\prod_{i=1}^3\left[\int_{-\pi}^{\pi}\! \frac{dk_i}{2\pi}\right]
  \frac{1}{z+\sum_{i=1}^{3}
  \left[
   \sin^2\left(\frac{k_i}{2}\right)
   +\frac{1}{3}\sin^4\left(\frac{k_i}{2}\right)
  \right]} 
\end{equation}
assuming $z>0$. The integral $I_0$ with five-point formula is obtained
by analytic continuation of the variable $z$. We are interested in the
coefficients of the first few terms of the expansion of $W(z)$ in
powers of $z$.

\comment{the method}

We employ the method similar to that Seki and van
Kolck~\cite{Seki:2005ns} used when they evaluated the Watson
integral. We start with the identity,
\begin{equation}
 \int_{0}^{\infty} d\alpha\, e^{-\alpha X} =\frac{1}{X}, \quad (X>0),
\end{equation}
and rewrite $W(z)$ as
\begin{equation}
 W(z) =\int_{0}^{\infty} d\alpha\, e^{-\alpha z}\left[U(\alpha)\right]^3,
 \label{walpha}
\end{equation}
where we have introduced 
\begin{equation}
  U(\alpha)=\int_{-\pi}^{\pi} \frac{dk}{2\pi}
 \exp\left\{
 -\alpha\left(
 \sin^2\left(\frac{k}{2}\right)
   +\frac{1}{3}\sin^4\left(\frac{k}{2}\right)
 \right)
 \right\}.
\end{equation}
By making a change of variable from $k$ to $t=\sin(k/2)$, we rewrite
it as
\begin{equation}
 U(\alpha) =\frac{2}{\pi}\int_{0}^{1}
  \frac{dt}{\sqrt{1-t^2}} e^{-\alpha(t^2+t^4/3)}.
  \label{ualpha}
\end{equation}

\comment{two regions}

We divide the integration region of $\alpha$ into two: the region
$0\le \alpha \le A$  and the region $A \le \alpha <\infty$. In the
former, we expand $e^{-\alpha z}$,
\begin{equation}
 \int_{0}^{A} d\alpha\, \left[U(\alpha)\right]^3
  -z\int_{0}^{A} d\alpha\, \alpha \left[U(\alpha)\right]^3+\cdots,
  \label{smallalpha}
\end{equation}
and evaluate the integrals numerically. In the latter, the dominant
contribution of the $U(\alpha)$ integral comes from the small $t$
region, so we approximate
\begin{equation}
 \frac{1}{\sqrt{1-t^2}}=e^{-\half\ln(1-t^2)}
  \approx e^{\half t^2+\frac{1}{4}t^4},
\end{equation}
and get 
\begin{align}
 U(\alpha)&\approx \frac{2}{\pi}\int_{0}^{\infty} dt\,
 \exp\left[-\left(\alpha-\half\right)t^2
 -\left(\frac{\alpha}{3}-\frac{1}{4}\right)t^4\right]
 \notag \\
 &=\frac{1}{\pi}\sqrt{\frac{a}{b}}e^{\frac{a^2}{2b}}
 K_{\frac{1}{4}}\left(\frac{a^2}{2b}\right),
 \label{ualphaapprox}
\end{align}
where $K_\nu(z)$ is the modified Bessel function and we have
introduced $a=\alpha-1/2$ and $b=4\alpha/3-1$. By using the asymptotic
expansion of $K_\nu(z)$,
\begin{align}
 K_{\nu}(z)&\sim \sqrt{\frac{\pi}{2z}}e^{-z}\sum_{n=0}^{\infty}
  \frac{\Gamma(\nu+n+\half)}{n!\Gamma(\nu-n+\half)(2z)^n},
 \notag \\
 &\qquad\qquad\qquad\qquad\quad (|\text{arg}z|<3\pi/2),
\end{align}
we obtain for a large value of $\alpha$ the following expansion,
\begin{equation}
 U(\alpha) =\frac{1}{\sqrt{\alpha\pi}}
 \left[
 1+\frac{1}{3}\left(\frac{1}{\alpha}\right)^2
 -\frac{35}{48}\left(\frac{1}{\alpha}\right)^3
 %+\frac{1225}{384}\left(\frac{1}{\alpha}\right)^4
 +\cdots
 \right].
\end{equation}

\comment{expansion of the second integral}

Substituting it into the integrand, we get the expansion
\begin{align}
 &\int_{A}^{\infty} d\alpha e^{-\alpha z}\left[U(\alpha)\right]^3
 \notag \\
  &=\frac{1}{A^{1/2}\pi^{3/2}}\int_{1}^{\infty}\frac{ds}{s^{3/2}}
 \, e^{-Azs}
  \bigg[
   1+\frac{1}{A^2}\left(\frac{1}{s}\right)^2
 \notag \\
 &{}\qquad\qquad\qquad\qquad\qquad\qquad\quad
 -\frac{35}{16A^3}\left(\frac{1}{s}\right)^3
   +\cdots
  \bigg]
 \notag \\
 &=\frac{1}{A^{1/2}\pi^{3/2}}
 \bigg\{
 \phi_{-\frac{3}{2}}(Az)
 +\frac{1}{A^2}\phi_{-\frac{7}{2}}(Az)
 \notag \\
 &{}\qquad\qquad\qquad
 -\frac{35}{16A^3}\phi_{-\frac{9}{2}}(Az)
 +\cdots
 \bigg\},
 \label{expressionwithphi}
\end{align}
where $\phi_m(z)$ is the incomplete Gamma function,
\begin{equation}
 \phi_m(z)=\int_{1}^{\infty} dt\, t^m e^{-zt}=z^{-(1+m)}\Gamma(1+m,z).
\end{equation}
Expanding Eq.~\eqref{expressionwithphi} in powers of $z$, we have
\begin{align}
 &\int_{A}^{\infty} d\alpha\, e^{-\alpha z}\left[U(\alpha)\right]^3
 \notag \\
 &=\frac{1}{A^{1/2}\pi^{3/2}}\left(2+\frac{2}{5A^2}-\frac{5}{8A^3}\right)
 \notag \\
 &{}\quad
 -\frac{2}{\pi}z^{1/2}
 \notag \\
 &{}\quad
 +\frac{1}{A^{1/2}\pi^{3/2}}
 \left(
 2A -\frac{2}{3A}+\frac{7}{8A^2}
 \right)z
 \notag \\
 &{}\quad
 +\frac{1}{A^{1/2}\pi^{3/2}}
 \left(
 -\frac{A^2}{3}+1 -\frac{35}{48A}
 \right)z^2
 \notag \\
 &{}\quad
 -\frac{8}{15\pi} z^{5/2} + \cdots.
 \label{largealpha}
\end{align}
Note that the coefficients of the half-odd-integer powers of $z$ do
not depend on $A$.

\comment{sum of two parts}

We numerically evaluate the sum of Eqs.~\eqref{smallalpha} and
\eqref{largealpha} for various values of $A$, we find that the sum is
independent of $A$ for a wide range of $A$. The coefficient of the
term of $\mathcal{O}(z^0)$ is about 0.876111 and that of
$\mathcal{O}(z^1)$ is about 0.106502, which correspond to
$\theta_1=1.37619\ldots$ and $R(0)=-0.412781\ldots$.

% Create the reference section using BibTeX:
\bibliography{LNEFT,NPRG,NEFT}

\end{document}